\documentclass[useAMS,usenatbib, usedcolumn]{mn2e}
\usepackage{aasmacros}
\usepackage{amssymb,amsmath}
\usepackage[pdftex]{graphicx}
\usepackage{url}
\usepackage{times}

\newcommand{\Sersic}{S\`{e}rsic}

\newcommand{\makebold}[1]{#1}

\title[Accretion and galaxy structure in CDM]{Galactic accretion and the outer structure of galaxies in the CDM model}
\author[Cooper et al.]{Andrew P. Cooper$^{1,2}$\thanks{E-mail:
acooper@nao.cas.cn}, Richard D'Souza$^{1}$, Guinevere Kauffmann$^{1}$, Jing Wang$^{1}$, \newauthor Michael Boylan-Kolchin$^{3}$, Qi Guo$^{2,4}$, Carlos S. Frenk$^{4}$ and Simon D.M. White$^{1}$\\
\\ $^{1}$Max-Planck-Institut f\"{u}r Astrophysik, 
Karl-Schwarzschild-Str. 1, D-85748, Garching, Germany\\
$^{2}$National Astronomical Observatories, Chinese Academy of Sciences, 20A Datun Road, Chaoyang, Beijing 100012, China\\
$^{3}$Center for Cosmology, Department of Physics and Astronomy, 4129 Reines Hall, University of California, Irvine, CA 92697, USA\\
$^{4}$Institute for Computational Cosmology, Department of Physics, University of Durham, South Road, Durham, DH1 3LE, UK}

\begin{document}

\date{Accepted xxxx. Received xxxx; in original form xxxx}

\pagerange{\pageref{firstpage}--\pageref{lastpage}} \pubyear{2013}

\maketitle

\label{firstpage}

\begin{abstract} We have combined the semi-analytic galaxy formation model of
  Guo et al. (2011) with the particle-tagging technique of Cooper et al.
  (2010) to predict galaxy surface brightness profiles in a representative
  sample of $\sim1900$ massive dark matter haloes ($10^{12}$--$10^{14}$
  $\mathrm{M_{\sun}}$) from the Millennium~II $\Lambda$CDM N-body simulation.
  Here we present our method and basic results focusing on the outer regions of
  galaxies, consisting of stars accreted in mergers. These simulations cover
  scales from the stellar haloes of Milky Way-like galaxies to the `cD
  envelopes' of groups and clusters, and resolve low surface brightness
  substructure such as tidal streams. We find that the surface density of
  accreted stellar mass around the central galaxies of dark matter haloes is
  well described by a \Sersic{} profile, the radial scale and amplitude of
  which vary systematically with halo mass ($M_{200}$). The total stellar mass
  surface density profile breaks at the radius where accreted stars start to
  dominate over stars formed in the galaxy itself. This break disappears with
  increasing $M_{200}$ because accreted stars contribute more of the total mass
  of galaxies, and is less distinct when the same galaxies are averaged in bins
  of stellar mass, because of scatter in the relation between $M_{\star}$ and
  $M_{200}$.  To test our model we have derived average stellar mass surface
  density profiles for massive galaxies at $z\approx0.08$ by stacking SDSS
  images.  Our model agrees well with these stacked profiles and with other
  data from the literature and makes predictions that can be more rigorously
  tested by future surveys that extend the analysis of the outer structure of
  galaxies to fainter isophotes. We conclude that it is likely that the outer
  structure of the spheroidal components of galaxies is largely determined by
  collisionless merging during their hierarchical assembly. 

\end{abstract}


\section{Introduction}

Hierarchical clustering leads to the coalescence of dark matter haloes.
Galaxies are formed `in situ' by the cooling and condensation of gas in the
cores of these haloes \nocite{WhiteRees78}({White} \& {Rees} 1978) and accrete
additional stars from the debris of their hierarchical progenitors. The aim of
this paper is to predict how the surface density profiles of galaxies reflect
changes in the balance between in situ star formation and stellar accretion
during their hierarchical growth over the lifetime of the universe.

The idea of using observations of accreted stars to test theories of
galaxy evolution has its roots in the study of the stellar halo and globular
clusters of the Milky Way and M31 \nocite{Baade44, ELS62, Searle78}({Baade}
1944; {Eggen}, {Lynden-Bell} \& {Sandage} 1962; {Searle} \& {Zinn} 1978).  The
recent discovery of cold stellar streams in these haloes, some with
identifiable progenitors, has provided direct evidence that they grow at least
partly through the tidal disruption of companion galaxies \nocite{Ibata95,
Belokurov06, NiedersteOstholt10, McConnachie09}(e.g. {Ibata}, {Gilmore} \&
{Irwin} 1995; {Belokurov} {et~al.} 2006; {Niederste-Ostholt} {et~al.} 2010;
{McConnachie} {et~al.} 2009).  Stellar haloes and streams appear to be a
generic feature of late-type galaxies \nocite{Zibetti04,
Richardson09,MartinezDelgado10, Bailin11, RadburnSmith11}({Zibetti}, {White} \&
{Brinkmann} 2004; {Richardson} {et~al.} 2009; {Mart{\'{\i}}nez-Delgado}
{et~al.} 2010a; {Bailin} {et~al.} 2011; {Radburn-Smith} {et~al.} 2011).
Shell-like structures have been detected around both early and late type
galaxies \nocite{Malin83, Schweizer80, Schweizer90, Schweizer92, Tal09,
MartinezDelgado10}({Malin} \& {Carter} 1983; {Schweizer} 1980; {Schweizer}
{et~al.} 1990; {Schweizer} \& {Seitzer} 1992; {Tal} {et~al.} 2009;
{Mart{\'{\i}}nez-Delgado}  {et~al.} 2010a) and can also be readily explained as
the result of galactic accretion in a cold dark matter (CDM) universe
\nocite{Cooper11b}(e.g {Cooper} {et~al.} 2011).  Galaxies at the centres of
massive clusters are often surrounded by extended envelopes of diffuse
`intracluster light' (ICL) \nocite{Matthews64, Oemler76, Thuan81, Schombert88,
Graham96, Lin04, Gonzalez05, Mihos05, Krick06, Donzelli11}({Matthews}, {Morgan}
\&  {Schmidt} 1964; {Oemler} 1976; {Thuan} \& {Romanishin} 1981; {Schombert}
1988; {Graham} {et~al.} 1996; {Lin} \& {Mohr} 2004; {Gonzalez}, {Zabludoff}, \&
{Zaritsky} 2005; {Mihos} {et~al.} 2005; {Krick}, {Bernstein}, \& {Pimbblet}
2006; {Donzelli}, {Muriel}, \&  {Madrid} 2011) which is also thought to
originate from the stripping and disruption of satellite galaxies
\nocite{Gallagher72}({Gallagher} \& {Ostriker} 1972).

Semi-analytic models of galaxy formation aim to quantify the importance of the
accretion of stars and gas in different types of galaxy \nocite{WhiteFrenk91,
Cole91, Kauffmann93, Somerville99, Cole00, Baugh05, DeLucia07,BensonBower10,
Guo11}({White} \& {Frenk} 1991; {Cole} 1991; {Kauffmann}, {White} \&
{Guiderdoni} 1993; {Somerville} \& {Primack} 1999; {Cole} {et~al.} 2000;
{Baugh} {et~al.} 2005; {De Lucia} \& {Blaizot} 2007; {Benson} \& {Bower} 2010;
{Guo} {et~al.} 2011). In particular, they predict how the mass of stars
accreted by a galaxy depends on the mass and assembly time of its dark matter
halo, as well as the number of progenitors of the halo and their individual
star formation histories. Such models predict that only the most
massive galaxies at the present day are dominated by stars accreted in mergers;
in galaxies less massive than the Milky Way, most stars form in situ from gas
cooling directly from their halo
\nocite{Baugh96,Kauffmann96,DeLucia06,Naab07,Purcell07,Guo08,Parry09}({Baugh},
{Cole} \& {Frenk} 1996; {Kauffmann} 1996; {De Lucia} {et~al.} 2006; {Naab}
{et~al.} 2007; {Purcell}, {Bullock}, \&  {Zentner} 2007; {Guo} \& {White} 2008;
{Parry}, {Eke}, \& {Frenk} 2009).

Accretion and merger events may still affect the colour, size and morphology of
a galaxy even if they make a limited contribution to its mass.  They are
therefore thought to be relevant to the dichotomy between early and late-type
morphologies in the Hubble sequence
\nocite{Toomre77,Fall79,Frenk85,Cowie94,Zepf97,Kauffmann98,Cole00,Bell04,Sales12}(e.g.
{Toomre} 1977; {Fall} 1979; {Frenk} {et~al.} 1985; {Cowie} {et~al.} 1994;
{Zepf} 1997; {Kauffmann} \& {Charlot} 1998; {Cole} {et~al.} 2000; {Bell}
{et~al.} 2004; {Sales} {et~al.} 2012) and many well-known scaling relations
between observable properties of massive galaxies
\nocite{FaberJackson76,Kormendy77,Djorgovski87,Peletier90,Bender92,Kauffmann03b,Bernardi03b}({Faber}
\& {Jackson} 1976; {Kormendy} 1977; {Djorgovski} \& {Davis} 1987; {Peletier}
{et~al.} 1990; {Bender}, {Burstein} \& {Faber} 1992; {Kauffmann} {et~al.}
2003b; {Bernardi} {et~al.} 2003), including correlations between the luminosity
of massive elliptical galaxies and the amplitude and shape of their projected
surface brightness profiles \nocite{Kormendy77, Binggeli84, Schombert86,
Graham03, Kormendy09, Graham11_arxiv}({Kormendy} 1977; {Binggeli}, {Sandage} \&
{Tarenghi} 1984; {Schombert} 1986; {Graham} \& {Guzm{\'a}n} 2003; {Kormendy}
{et~al.} 2009; {Graham} 2011, and references therein).

The effects of stellar accretion on galactic structure depend on the population
of infalling galaxies and the rate at which haloes coalesce as well as the
gravitational dynamics of the accretion process.  This means that galactic
accretion cannot be studied in isolation from galaxy formation. The way in
which accreted stars are deposited in the outer regions of galaxies and the
consequent change in observables such as half-light radius and stellar mass has
been considered extensively in recent literature \nocite{Daddi05, Trujillo06,
Van-Dokkum10}({Daddi} {et~al.} 2005; {Trujillo} {et~al.} 2006; {van Dokkum}
{et~al.} 2010).  Simulations focussing on this issue by \nocite{Naab06a,
Naab07, Naab09b}e.g. {Naab}, {Khochfar} \& {Burkert} (2006); {Naab} {et~al.}
(2007); {Naab}, {Johansson}, \& {Ostriker} (2009) and \nocite{Oser10}{Oser}
{et~al.} (2010) have highlighted the importance of N-body dynamical simulations
when making quantitative predictions for the evolution of galaxy sizes and
velocity dispersions in CDM \nocite{Gonzalez09, Guo11, Shankar13}(compare
{Gonz{\'a}lez} {et~al.} 2009; {Guo} {et~al.} 2011; {Shankar} {et~al.} 2013).
Notably, these simulations suggest that high mass ratio mergers contribute
significantly to the structure of massive elliptical galaxies
\nocite{Hilz12,Hilz13}({Hilz} {et~al.} 2012; {Hilz}, {Naab} \&  {Ostriker}
2013). 

In this paper we use an extension of the \nocite{Guo11}{Guo} {et~al.} (2011,
hereafter G11) semi-analytic galaxy formation model to predict the spatial
distribution of all stars accreted on to present-day massive galaxies. Like
many recent numerical studies of the size evolution of massive quiescent
galaxies \nocite{Meza03, Naab09b, Oser10}({Meza} {et~al.} 2003; {Naab} {et~al.}
2009; {Oser} {et~al.} 2010) and Milky Way-like galaxies
\nocite{Abadi06,Cooper10,Font11,Sales12}({Abadi}, {Navarro} \& {Steinmetz}
2006; {Cooper} {et~al.} 2010; {Font} {et~al.} 2011; {Sales} {et~al.} 2012) we
emphasize the difference between in situ star formation and galactic accretion
in our analysis.  The semi-analytic component of our model provides the full in
situ star formation histories of galaxies and all their hierarchical
progenitors, matching constraints such as the galaxy stellar mass function. In
addition, we use an N-body method to predict how each accreted population
evolves in all six dimensions of phase space. Our model is applied to the
Millennium II simulation, which contain $\sim2000$ haloes in the mass range
$10^{12} <M_{200} < 10^{14}\, \mathrm{M_{\odot}}$. We can therefore make a
statistical comparison to observational data on the total amount of accreted
light around galaxies and its distribution. In this paper, we only consider
galaxies at the centres of virialised dark haloes at $z=0$, although our model
can also make predictions for galaxies at high redshift and for galaxies that
are satellites at the present day.  We focus on the spatial distribution of
accreted stars, but we note that our model also predicts their kinematic
properties and chemical abundances. 

A statistical study of this sort is motivated by the availability of moderately
deep wide-field imaging from surveys such as SDSS Stripe 82 and PanSTARRS,
which will enable us to determine the extent to which results obtained from the
Milky Way and M31 stellar haloes are applicable to the galaxy population as a
whole.  Previous studies of galaxy structure in large surveys have focused on
regions of high surface brightness \nocite{Shen03}(e.g. {Shen} {et~al.} 2003),
because the outskirts of galaxies are usually much fainter than the sky, even
in the case of the envelopes around brightest cluster galaxies (BCGs).
Stacking \nocite{Zibetti05, Tal11}(e.g. {Zibetti} {et~al.} 2005; {Tal} \& {van
Dokkum} 2011) is a promising technique for studying the average properties of
these regions.  We have therefore carried out our own stacking analysis using
imaging data from SDSS, for comparison to our models. 

We proceed as follows. In Section~\ref{sec:simulations} we describe how we
select a sample of massive central galaxies from the Millennium II simulation.
We also summarize how our particle-tagging method works.
Section~\ref{sec:individual} shows examples of the stellar haloes of individual
galaxies in our model. In Section~\ref{sec:average_profiles} we present our
main statistical results in the form of average stellar mass surface density
profiles for haloes and galaxies of different mass, highlighting differences
between in situ and accreted stars.  We compare with observations from the
literature in Section~\ref{sec:obs} and from our own stacking analysis in
Section~\ref{sec:sdss}.  Section~\ref{sec:origin} interprets trends in surface
brightness profile shape by studying the origin of accreted stars in our
simulation.  We summarise our findings in Section~\ref{sec:conclusions}.
Appendix~\ref{appendix_b} discusses the numerical convergence of our method and
differences with our previous work on particle tagging.
Appendix~\ref{appendix_d} describes the technical details of our SDSS stacking
analysis.  

\section{Simulations}
\label{sec:simulations}

\subsection{Semi-analytic model}

\begin{figure}
  \includegraphics[width=84mm, trim=0cm 1cm 0cm 0.5cm, clip=True]{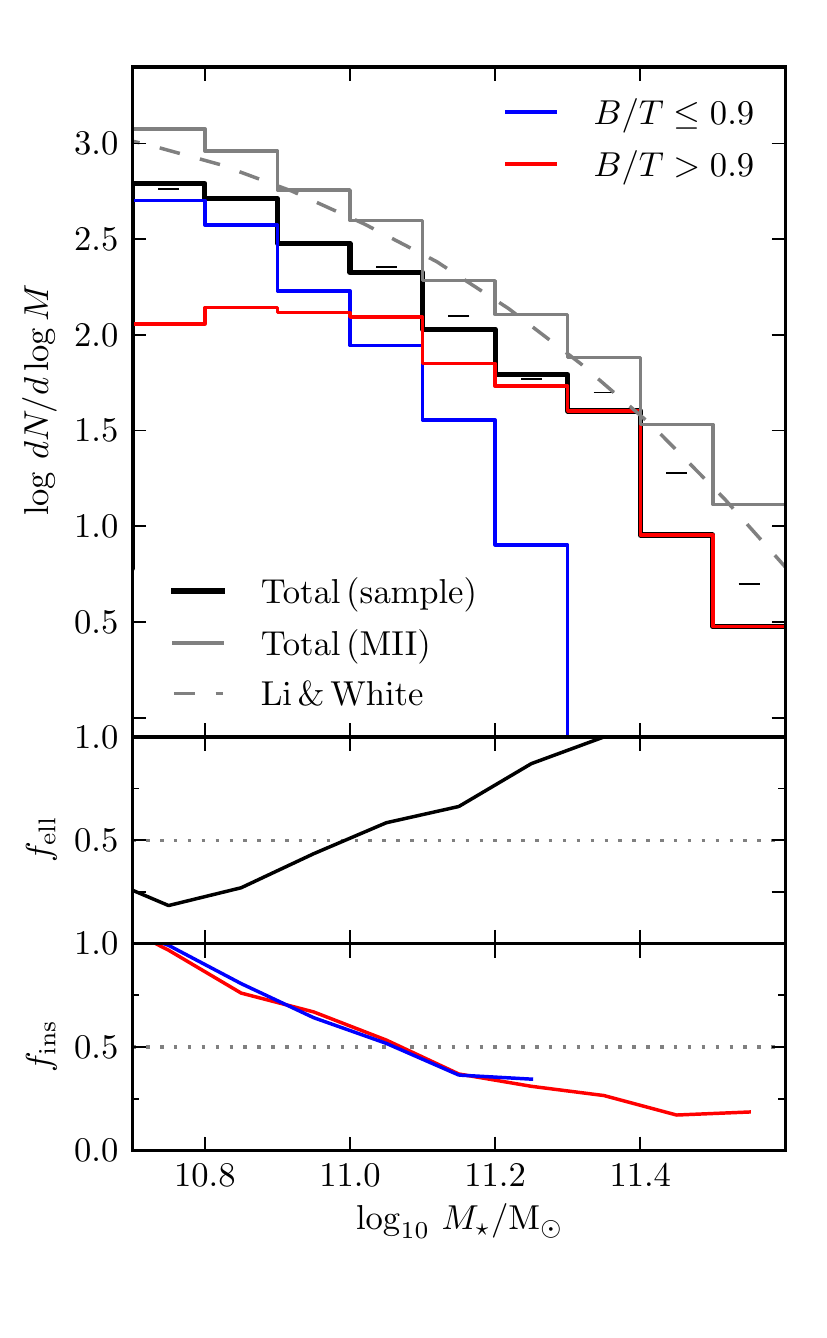}

  \caption{The solid black histogram shows the mass function of our primary
  sample of central galaxies from the G11 model. The solid grey histogram is
  the mass function of \textit{all} galaxies in the Millennium~II simulation,
  including satellite galaxies and all galaxies associated with the merger tree
  of the most massive cluster which we have excluded from our analysis.  The
  dashed grey line shows the Li \& White (2009) SDSS mass function normalized
  to the Millennium~II volume, derived as in Appendix A of Guo~et~al. (2010)
  and assuming the same Chabrier IMF as our model.  Red and blue histograms
  split the mass function of our primary sample into two components based on
  bulge-to-total mass ratio, as shown in the legend. Short black horizontal
  lines (between the black and grey histograms) show the effect of including
  the G11 prediction of diffuse stellar halo mass for each galaxy in our
  sample. Considering only our sample galaxies, the lower two panels show (top)
  the fraction, $f_{\mathrm{ell}}$, of galaxies in each mass bin with $B/T >
  0.9$, and (bottom) the fraction of the combined bulge and disc stellar mass
  in each bin that was formed in situ, split by $B/T$ as in the main panel.} 

  \label{fig:mass_function_sample} 
\end{figure}

Millennium~II \nocite{BoylanKolchin09}({Boylan-Kolchin} {et~al.} 2009) is a
collisionless $N$-body simulation of $\Lambda$CDM structure formation in a
comoving volume of $10^{6} \, h^{-3} \, \mathrm{Mpc^{3}}$ with a flat
$\Lambda$CDM cosmology, $\Omega_{\mathrm{m}}=0.25$,
$\Omega_{\mathrm{\Lambda}}=0.75$ and Hubble parameter $h=0.73$. The particle
mass is $6.89 \times 10^6\, h^{-1} \, \mathrm{M_{\odot}}$. The semi-analytic
galaxy model of G11 is based on halo merger trees derived from Millennium~II
and its parameters are tuned to fit the SDSS stellar mass function of
\nocite{Li09massfunction} Li \& White (2009; grey dashed line in
Fig.~\ref{fig:mass_function_sample}).  Figure~7 of G11 shows that when same
model is applied to the Millennium Simulation \nocite{Springel05MS}({Springel}
{et~al.} 2005), which has a larger volume, it overpredicts the number of
galaxies with $\log_{10} M_{\star}/\mathrm{M_{\sun}} \gtrsim 11.5$.  G11
suggest that this discrepancy is due to sample variance and $\sim
1\,\mathrm{mag}$ luminosity uncertainties for the most luminous galaxies in
SDSS (see e.g. \nocite{Bernardi13_arxiv} Bernardi et al. 2013).  

From the results of G11 we select 1872 {\em central} galaxies more massive than
$M_{\star}=5\times10^{10}\,\mathrm{M_{\sun}}$ at $z=0$.  The stellar mass
function of these galaxies is shown by the black histogram in
Fig.~\ref{fig:mass_function_sample}. We do not select all the galaxies in the
simulation above our threshold stellar mass (in particular, we exclude
satellite galaxies and the central galaxy of the most massive cluster). The
grey histogram shows the mass function of all galaxies in the simulation to
demonstrate that this selection does not bias our sample.

The G11 model uses a combination of two 1D axisymmetric density profiles to
represent the distribution of stars inside galaxies (an exponential disc and a
Jaffe-model bulge\footnote{G11 also track the total mass in a diffuse
stellar halo component but do not specify its density profile. The mass in this
component is only a significant fraction of the total galaxy stellar mass above
$\log_{10} M_{\star}/ \mathrm{M_{\odot}} \sim 11.2$. Horizontal bars in
Fig.~\ref{fig:mass_function_sample} show how including the G11 stellar halo
component in the central galaxy stellar mass affects the stellar mass function
of our sample at $z=0$.}) and quantifies galaxy morphology using the
ratio of bulge mass to total stellar mass, $B/T$. The red histogram in
Fig.~\ref{fig:mass_function_sample} shows the mass function of galaxies with
$B/T \geq 0.9$ (`ellipticals') and the blue histogram the mass function of
galaxies with $B/T<0.9$. We plot the ratio of these mass functions in the
middle panel. The fraction of `elliptical' galaxies increases from 50 to 100
per cent in the interval $11.0 <\log_{10} \mathrm{M_{\star}/M_{\sun}} < 11.3$.
This is in good agreement with observations \nocite{Conselice06_classify}(e.g.
{Conselice} 2006, see figure~4 of G11) and a similar transition scale has been
found in other semi-analytic models \nocite{DeLucia11}({De Lucia} {et~al.}
2011). These models also predict differences between early and late-type K-band
luminosity functions that are qualitatively similar to those observed
\nocite{Benson10}({Benson} \& {Devereux} 2010).

We define `in situ' stars as those that are still gravitationally bound to the
dark matter halo in which they formed. G11 predict that more massive galaxies
form less of their total stellar mass in situ. The lowest panel of
Fig.~\ref{fig:mass_function_sample} shows that the fraction of stars in each
mass bin formed in situ decreases from almost (but of course not exactly) 100
per cent at $\log_{10} \mathrm{M_{\star}/M_{\sun}} = 10.7$ to 50 per cent at
$\log_{10} \mathrm{M_{\star}/M_{\sun}} = 11.1$. The mass fraction of in situ
stars in systems with $B/T \geq 0.9$ is similar to that of other galaxies
regardless of stellar mass. This implies that in the G11 model, the relative
contribution of in situ star formation depends primarily on stellar mass and
not morphology\footnote{Since galaxies with $B/T > 0.9$ were formed by low
mass ratio mergers, one might expect their in situ fractions to be lower
than those of galaxies with the same mass having $B/T < 0.9$. However, in the G11
model, merger-induced starbursts and disc instabilities can increase the in
situ mass while also increasing $B/T$.}. For the most massive galaxies plotted
Fig.~\ref{fig:mass_function_sample}, the fraction of stars formed in situ is
$\sim$19 per cent. Thus the model makes a clear prediction that accreted
stellar populations will dominate the structure of the most massive galaxies. 

\subsection{Particle tagging}

We use a technique we call particle tagging to predict the stellar population
mix and spatial distribution of stars in galaxies, based on the merger trees
and star formation histories of the G11 model. This technique uses additional
information from the underlying N-body simulation in order to predict more
observables than standard semi-analytic models, without running a new
simulation. Other studies using particle tagging techniques
include \nocite{Bullock01_BKW}{Bullock}, {Kravtsov} \&  {Weinberg} (2001),
\nocite{Napolitano03}{Napolitano} et al. (2003),
\nocite{Bullock05}{Bullock} \& {Johnston} (2005),
\nocite{Penarrubia08_tidal_sim}{Pe{\~n}arrubia}, {Navarro} \&  {McConnachie}
(2008)  and \nocite{Laporte13_arxiv}{Laporte} et al. (2013), although these
were not coupled to the predictions of semi-analytic models. We give a brief
summary of our method below; for more detail see \nocite{Cooper10}{Cooper}
{et~al.} (2010, hereafter C10).  A discussion of minor differences between our
implementation and that of C10 and a test of convergence with their results can
be found in Appendix \ref{appendix_b}.

The particle tagging technique associates (`tags') sets of dark matter
particles in an N-body simulation (here Millennium~II) with stellar populations
of a single metallicity and age. The tagged particles can be used to track the
evolution of their associated population in phase space, from the time when the
stars form to the present day ($z=0$). Our definition of a stellar population
comprises all the stars formed in a single galaxy between two consecutive
snapshots of the G11 model. An isolated galaxy that forms stars at a constant
rate for a Hubble time will produce a number of these populations equal to the
number of simulation snapshots. All model galaxies at $z=0$ are a superposition
of many such populations, because they accrete populations formed in their
hierarchical progenitors as well as forming their own stars in situ.

For every population, particles are selected according to a tagging criterion
(described below). An equal fraction of the total mass of the population is
given to each particle thus selected.  Every new population tags a new  set of
particles, selected from the corresponding dark matter halo at the snapshot
immediately after the population forms. This means that a DM particle can be
tagged more than once, if it meets the tagging criterion for two or more
populations (by construction, this can only happen at different snapshots).  In
such cases, each tag is tracked separately. A corollary is that each tagged
particle carries its own unique star formation and enrichment history, with the
time resolution of the Millennium~II snapshots.

\subsection{Tagging criterion and the $f_{\mathrm{mb}}$ parameter}
\label{sec:fmb}

The particles we select for tagging are supposed to approximate the phase space
distribution of the stars immediately after they form. Stars are the end result of dissipative
collapse, so a basic requirement is that particles tagged with newly-formed
stars should be deeply embedded in the potential well of their dark halo when
we tag them. We achieve this by ranking DM particles in the halo by their
binding energy and selecting all those more bound than a threshold value,
corresponding to a fixed fraction of the mass of the halo. Following C10, we
call this free parameter of the method the `most-bound fraction',
$f_{\mathrm{mb}}$. A value of $f_{\mathrm{mb}}=0.01$ means we selected the 1
per cent most-bound particles.

\begin{figure} 
  \includegraphics[width=84mm, clip=True]{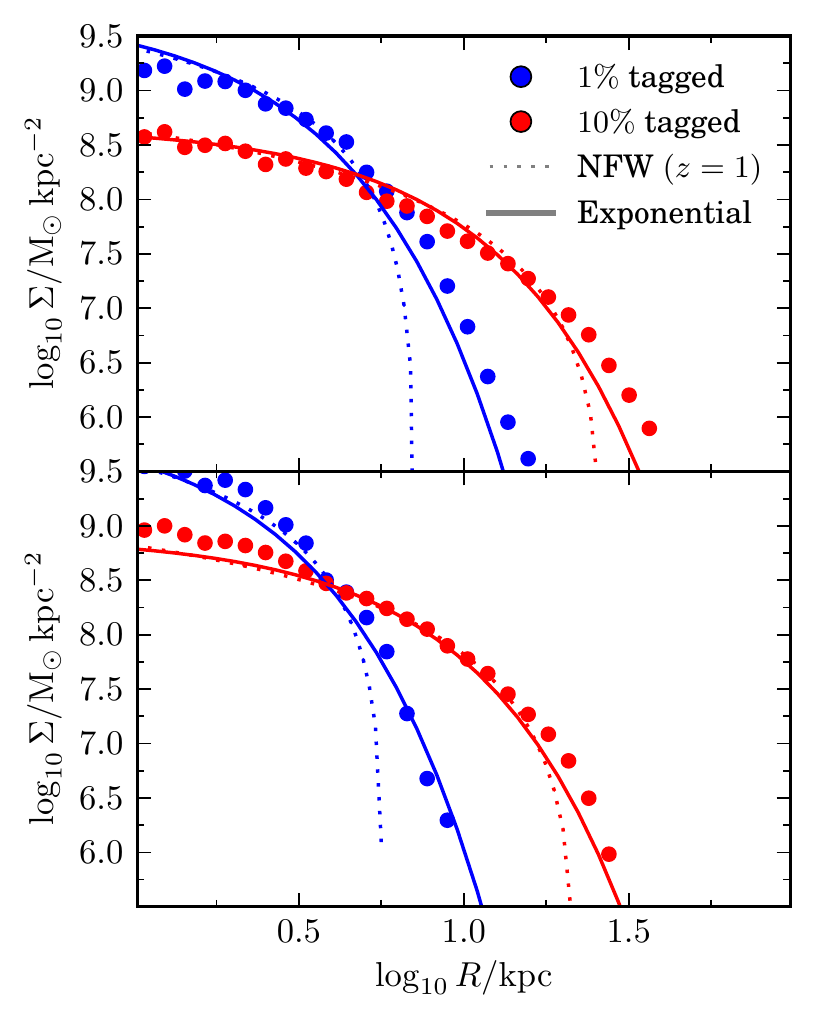}

  \caption{Dots show in situ surface density profiles in two Milky Way-like
  haloes at $z=0$ from G11, predicted by our particle tagging model with
  $f_{\mathrm{mb}} = 1\%$ (blue) and $10\%$ (red). Upper and lower panels
  respectively correspond to galaxies with $M_{200}=(12.1,12.3)$,
  $M_{\star}=(10.8,10.9)$ and NFW concentration $c=(7.2, 8.2)$. Dotted lines show
  density profiles for the corresponding fractions of most bound DM particles at
  $z=1$ (assuming an isotropic NFW distribution function with virial radius and
  concentration given by the N-body halo of each galaxy), normalized to the same
  stellar mass. Solid lines show exponential profiles with the same amplitude 
  and half mass radius as the dotted lines.} 

  \label{fig:isexamples}
\end{figure}

The choice of $f_{\mathrm{mb}}$ is more-or-less arbitrary, but this freedom
allows us to tune the scale length of the in situ components of our galaxies in
a predictable way. This is because, in an NFW  potential
\nocite{NFW96}(Navarro, Frenk \& White 1996), the surface density profile of
dark matter more bound than a given energy is roughly exponential (at least for
$f_{\mathrm{mb}}< 10$ per cent), with a scale radius that depends on the
threshold energy. This result can be verified easily by integrating the
cumulative energy distribution of an NFW halo up to a given fraction of its
virial mass, and constructing the corresponding density profile from the phase
space distribution function. We have done this using numerical approximations
for the distribution function and density of states given by
\nocite{Widrow00}{Widrow} (2000) for a spherical NFW halo with an isotropic
velocity distribution. 

To illustrate this point, Fig.~\ref{fig:isexamples} shows the profile of in
situ stars in two `Milky Way' mass haloes from Millennium~II (top and bottom
panels), according to our full particle tagging model (dots) with
$f_{\mathrm{mb}} = 1\%$ (blue) and $f_{\mathrm{mb}} = 10\%$ (red). Dotted lines
show the profile we obtain using the \nocite{Widrow00}{Widrow} (2000)
distribution function to select the equivalent most-bound mass fraction at
$z=1$, by which time most of the stars in these galaxies have already formed
(the central regions of these haloes are very stable thereafter, e.g.
\nocite{Wang11}{Wang} {et~al.} 2011). The dotted profiles are not exactly
exponential because our procedure obviously imposes an energy threshold, which
corresponds to a truncation radius.  Solid lines show exponential profiles that
have the same scale radius as the dotted profiles -- these roughly approximate
the diffusion of tagged particles across the initial energy threshold over
time. Note that because we perform our tagging procedure at every snapshot,
each new population in our full model will have a different amount of time to
diffuse away from its initial configuration. 

We stress that our model for the structure of merger remnants is not purely
collisionless, because the G11 model explicitly includes enhanced dissipative
star formation (in the bulge component) during mergers. This is important
because hydrodynamical simulations of galaxy mergers have shown that nuclear
starbursts increase the central phase space density of merger remnants
\nocite{Hernquist93_phasespace,Robertson06,Hopkins08}({Hernquist}, {Spergel} \&
{Heyl} 1993; {Robertson} {et~al.} 2006; {Hopkins} {et~al.} 2008). We include
stars formed in these bursts in our tagging in the same way as those formed in
the `quiescent' mode.

\subsection{Constraints on $f_{\mathrm{mb}}$ from the galaxy mass--size relation}
\label{sec:size_mass}

\begin{figure}
  \includegraphics[width=84mm,clip=True]{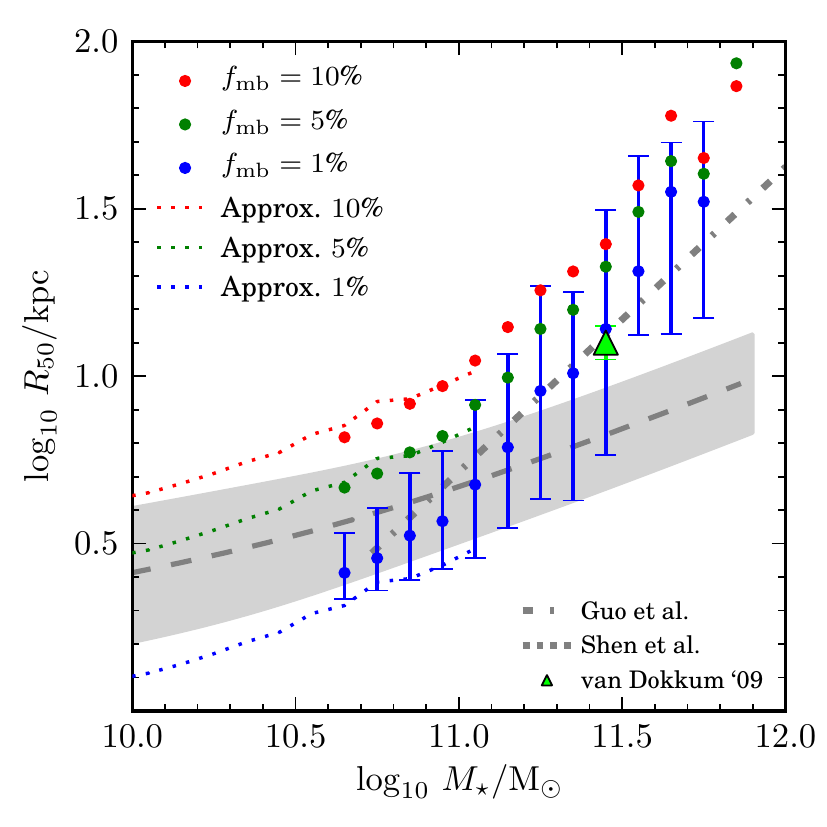}

  \caption{Points plot stellar mass $M_{\star}$ against projected half-mass
  radius $R_{50}$ for all galaxies in our full particle tagging models; colours
  correspond to different values of $f_{\mathrm{mb}}$.  Dotted lines of the
  same colours show our analytic approximation for the sizes of all central
  galaxies in Millennium II, using their in situ stellar mass and $z=1$ halo
  properties.  The dashed grey line and shaded region plot the late-type galaxy
  relation of Shen {et~al.} (2003) and its $1\sigma$ range. The dot-dashed grey
  line shows the early-type galaxy relation of {Guo} et~al. (2009) corrected
  from a Kroupa to a Chabrier IMF using $\Delta \log_{10} M_{\star}/M_{\odot} =
  -0.04$. The green triangle corresponds to a stack of deep images of 14 nearby
  ellipticals ({van Dokkum} {et~al.} 2010).}
  \label{fig:size_mass_all} 
\end{figure}

From the above we conclude that our analytic approximation can reproduce the
$z=0$ density profiles of particles that we tag to represent in situ stars with
reasonable accuracy. This provides an intuitive understanding of why the
parameter $f_{\mathrm{mb}}$ sets the sizes of galaxies dominated by in situ
star formation, and how those sizes vary with the scale of the dark matter
potential. Based on this, we can determine a range of suitable
$f_{\mathrm{mb}}$ values empirically, by comparing the predicted relation
between stellar mass ($M_{\star}$) and projected half-mass radius ($R_{50}$) to
observations of galaxies dominated by in situ star formation, i.e. those with
$\log_{10} M_{\star} \lesssim 10.8$; e.g. \nocite{Guo08} Guo \& White 2008).
By using only in situ stars to constrain $f_{\mathrm{mb}}$, the distribution of
accreted stars remains a valid prediction of our model. With a similar
approach, C10 found that $f_{\mathrm{mb}}=1\%$ gave reasonable agreement
between their simulations and the $M_{\star}$--$R_{50}$ relation of dwarf
satellite galaxies in the Local Group. We re-examine the choice of
$f_{\mathrm{mb}}$ because C10 considered only galaxies that were predominantly
satellites, with very different stellar and dark matter mass scales to those in
our simulation.

In Fig.~\ref{fig:size_mass_all} we show the median $M_{\star}$--$R_{50}$
relation for all galaxies in our full particle tagging model at $z=0$ (red,
green and blue points, corresponding to $f_{\mathrm{mb}}=10\%, 5\%$ and $1\%$
respectively). Each galaxy in the model contributes three values of $R_{50}$
from three orthogonal projections to its $M_{\star}$ bin. Model galaxies have a
considerable scatter in $R_{50}$ at fixed stellar mass (the $16$--$84$
percentile range for $f_{\mathrm{mb}}=1$ per cent is shown by the error bars on
the blue points; other values of $f_{\mathrm{mb}}$ have very similar scatter).
Note that our sample contains only 13 galaxies with $M_{\star} >
10^{11.5}\mathrm{M_{\odot}}$.

At $M_{\star}\lesssim10^{11} M_{\odot}$, where galaxies in our model are
dominated by in situ stars, we can use our analytic approximation to
extrapolate our results below the limit of $M_{\star} \ge 5\times10^{10}
\mathrm{M_{\odot}}$ we imposed when selecting galaxy merger trees for tagging.
We take all central galaxies at $z=0$ in the G11 Millennium II catalogue with
$10^{8} \le M_{\star} < 10^{11.1} \mathrm{M_{\odot}}$ and use the virial radius
($r_{200}$) and concentration\footnote{We determine $c$ as $c = 2.16 \,
r_{200}/r_{\mathrm{max}}$ where $r_{\mathrm{max}}$ is the radius of maximum
circular velocity \nocite{Cole96}(e.g {Cole} \& {Lacey} 1996).} $c$ of their
dark matter haloes at $z=1$ to predict $R_{50}$ at $z=0$.
These predictions are shown as dotted lines in
Fig.~\ref{fig:size_mass_all}. They agree roughly with the results of the full
tagging model in the range where they overlap, although we find that $R_{50}$
is underestimated by $\sim33\%$ for $f_{\mathrm{mb}}=1\%$.  This is not
surprising as the $f_{\mathrm{mb}}=1\%$ results are most sensitive to the
simple representation of the central dark matter potential.

\makebold{This extrapolation confirms that our tagging model produces a curved $\log_{10}
M_{\star}$--$\log_{10} R_{50}$ relation similar to the observed relation of
\nocite{Shen03}{Shen} {et~al.} (2003) for late-type galaxies (grey dashed line
with $1\sigma$ scatter).  At lower $M_{\star}$ the model with
$f_{\mathrm{mb}}=1\%$ underpredicts the observed relation. Our approximation
predicts that a model with $f_{\mathrm{mb}}\approx3\%$ would be closer to the
data; $f_{\mathrm{mb}}=5\%$ is also plausible, as the \nocite{Shen03}{Shen}
{et~al.} (2003) relation may underpredict the sizes of edge-on galaxies by
$\sim0.15$~dex \nocite{Dutton07}(e.g. {Dutton} {et~al.} 2007). On the other
hand, the relation from our model also includes early-type galaxies, which are
known to be significantly more compact at than late-types at $M_{\star}
\lesssim 10^{11} \mathrm{M_{\odot}}$ (\nocite{Shen03}{Shen} {et~al.} 2003).
\nocite{Kauffmann03a}{Kauffmann} {et~al.} (2003a) do not separate galaxies by
morphology and find $ R_{50} = 2.38 \, \mathrm{kpc}$ ($h=0.73$) for $10.0 <
\log_{10} M_{\star} < 10.5$ which supports $1\% \lesssim f_{\mathrm{mb}}
\lesssim 3\%$.}

\makebold{Having determined a plausible range of $f_{\mathrm{mb}}$ with reference to in
situ stars, we can now examine the predictions of these models for more massive
galaxies that are dominated by accreted stars. In Fig.~\ref{fig:size_mass_all},
the mass--size relation clearly steepens at $M_{\star} > 10^{11}
\mathrm{M_{\odot}}$  (e.g. \nocite{Shen03}{Shen} {et~al.} 2003;
\nocite{Hyde09}{Hyde} \& {Bernardi} 2009). In this regime the
$f_{\mathrm{mb}}=1\%$ model follows approximately the relation for early-type
galaxies found by \nocite{Guo09}{Guo} {et~al.} (2009; grey dot-dashed
line\footnote{We note that the observations plotted in figure 8 of
\nocite{Guo09}{Guo} {et~al.} (2009) for $M_{\star} >
10^{11.5}\mathrm{M_{\odot}}$ lie systematically above the linear relation we
plot, which suggests that the relation may curve upwards at higher $M_{\star}$
(e.g.  \nocite{Bernardi11_mnmain}{Bernardi} {et~al.} 2011).}). This relation
agrees with the results of \nocite{Van-Dokkum10}{van Dokkum} {et~al.} (2010)
(green triangle), who stacked \Sersic{} fits to individual deep images of 14
galaxies with $\log_{10} \langle M_{\star} \rangle /M_{\odot}=11.45$ from a
mass-selected and approximately volume-limited sample of early types
(\nocite{Tal09}{Tal} {et~al.} 2009).}

\makebold{A value of $f_{\mathrm{mb}}=5\%$, which gives a reasonable scale for in
situ stars in lower-mass galaxies, overpredicts $R_{50}$ from Guo {et~al.}
(2009) by $\sim0.15$~dex at $M_{\star} > 10^{11} \, \mathrm{M_{\odot}}$. A
value of $f_{\mathrm{mb}}=10\%$ overpredicts $R_{50}$ by $\lesssim0.3$~dex,
indicating that our predictions for $R_{50}$ in this mass range are less
sensitive to $f_{\mathrm{mb}}$ than they are for in-situ dominated galaxies,
where $R_{50}$ changes by $\sim0.5$~dex over the same range in
$f_{\mathrm{mb}}$.  As we will show in the following section, this
sensitivity is still mostly driven by the strong effect of
$f_{\mathrm{mb}}$ on the scale of in situ stars even in very massive galaxies.
The effects on the accreted component are even weaker, the most notable being
an increase in $R_{50}$ with $f_{\mathrm{mb}}$ because more extended satellite
galaxies are more easily stripped.}

\makebold{The apparent excess in $R_{50}$ at $M_{\star}>10^{11.5} \, \mathrm{M_{\odot}}$
even for an $f_{\mathrm{mb}}\sim1\%$ model may arise in part because masses and
sizes derived from SDSS photometric measurements miss a substantial fraction of
light in the outer regions of massive galaxies with high \Sersic{} index
\nocite{Graham05_petro, Bernardi07, Lauer07, Blanton11, Bernardi12_arxiv,
Meert12_arxiv, Mosleh13_arxiv} ({Graham} {et~al.} 2005; {Bernardi} {et al.}
2007; {Lauer} {et~al.} 2007; {Blanton} {et~al.} 2011; {Bernardi} {et~al.} 2012;
{Meert} {et~al.} 2012; {Mosleh}, {Williams} \& {Franx} 2013).}

\makebold{Subject to this uncertainty, we conclude that the median half-mass radii of
accretion-dominated galaxies in our models are consistent with observations
for $1\% < f_{\mathrm{mb}} \lesssim 3\%$. We prefer not to fine-tune a `best'
choice of $f_{\mathrm{mb}}$. Effects including the treatment of in situ star
formation in G11, the cosmology of Millennium II and the accuracy of observed
$M_{\star}$ and $R_{50}$ values (e.g.  \nocite{Bernardi13_arxiv,
Mitchell13_arxiv} Bernardi et al.  2013, Mitchell et al. 2013 and references
therein) may determine how well our tagging model matches the observations,
because $1\%$ changes in $f_{\mathrm{mb}}$ correspond to $0.1$~dex differences
in $R_{50}$ and $M_{\star}$.   Therefore, in order to bracket the plausible
range of $f_\mathrm{mb}$ and to show how it affects our conclusions, we will
use fiducial values of $f_{\mathrm{mb}}=1\%, 5\%$ and $10\%$ for the remainder
of the paper.}

\subsection{Limitations of the method}
\label{sec:limitations}

The most important limitation of particle tagging is that it neglects the
gravitational effects of concentrating baryons in the cores of dark matter
haloes. C10 considered only the highly dark matter dominated dwarf
satellites of Milky Way-like galaxies. Here, however, we tag dark matter
particles in galaxies of Milky Way mass and larger, where the
\makebold{gravitational potential within $R_{50}$} could be modified significantly
by baryons \makebold{(e.g. \nocite{Koopmans09}{Koopmans et al.  2009}). If this
effect was included in our model, our central galaxies may be more concentrated
and the cores of our satellites more robust to tidal stripping.  These effects
may explain the $\sim0.1$~dex overestimate of $R_{50}$ at
$M_{\star}\gtrsim10^{11} \mathrm{M_{\odot}}$ in Fig.~\ref{fig:size_mass_all}
(for an $f_{\mathrm{mb}}\sim3\%$ model)}. Another limitation is that a tagging
method based on binding energy alone cannot model rotationally supported discs,
thus even `late type' galaxies are represented by dispersion-supported
spheroidal systems\footnote{We can still separate galaxies by morphology using
the G11 $B/T$ ratio, which is independent of our tagging approach.}. Finally,
we neglect the possibility that in situ stars form on very loosely bound orbits
far away from galaxies, for example from cold gas clumps stripped from
satellites or ejected in galactic fountains. 

The first two limitations mean that, in the N-body part of our model, we cannot
include the formation and secular evolution of discs, adiabatic contraction or
the destruction of cusps in dark haloes by feedback \nocite{Navarro96,
Gnedin04, Dutton07, Pontzen12}({Navarro}, {Eke} \& {Frenk} 1996; {Gnedin}
{et~al.} 2004; {Dutton} {et~al.} 2007; Pontzen \& Governato 2012). As discussed
in the previous section, the stellar mass surface density profiles of
late-type galaxies in our model are exponential, and their scale lengths can
be roughly matched to observations by choosing $f_{\mathrm{mb}}$ appropriately.
\makebold{Nevertheless, the energy and angular momentum distribution of stars in
disc galaxies may affect the details of their stripping and disruption (e.g.
\nocite{Chang13} {Chang, {Macci\`{o}} \& Kang} 2013)}. Satellites in our model
do not suffer tidal shocking or enhanced (stellar) dynamical friction from the
stars of their central galaxy, which may artificially favour their survival.
\makebold{On the other hand, both limited numerical resolution and the absence of
gravitational binding energy from stars will artificially enhance the rate of
satellite disruption}.

We proceed with these approximations nevertheless as a first step towards
modelling the spatial distribution and other properties of the diffuse stellar
component of galaxies, in the context of a realistic model of galaxy formation.
Our model for in situ stars is only intended to serve as a means of creating
initial conditions for accretion events with roughly the right scale and
concentration, and as a means of quantifying the relative contribution of
accreted stars at different radii.

\begin{figure*}
  \includegraphics[width=140mm, clip=True, trim=1cm 1.2cm 1cm 1.2cm]{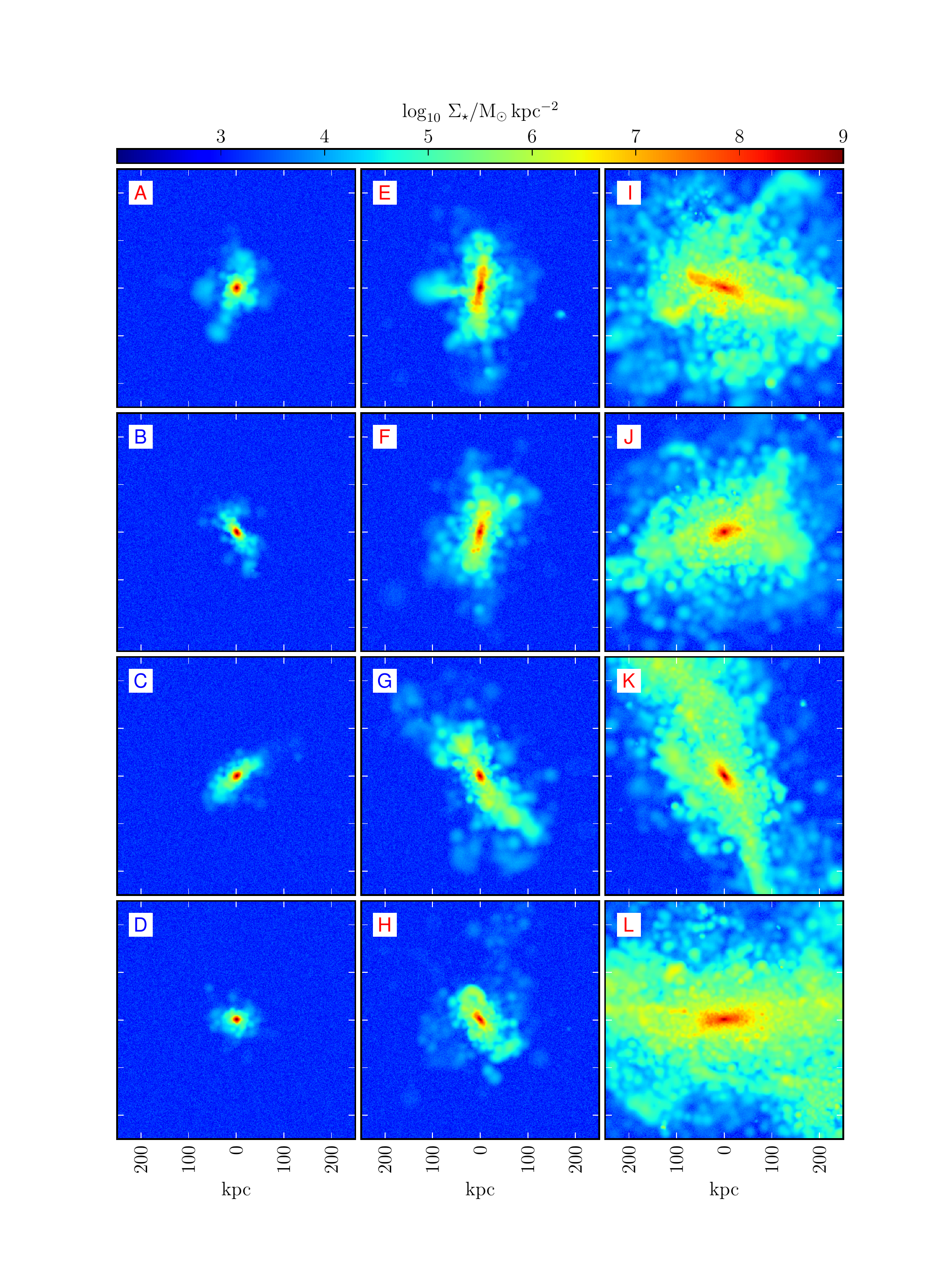}

  \caption{Projected stellar mass surface density in regions $500\times500$~kpc
  around 12 simulated galaxies. In each of the three columns, $M_{200}$ and
  $M_{\star}$ are approximately constant, representing (from left to right) Milky
  Way-like galaxies, massive isolated galaxies, and group/cluster central
  galaxies. These examples are referred to in the text by their labels A--L (blue
  where G11 $B/T \le 0.9$ and red where G11 $B/T > 0.9$). More details are shown
  in Fig.~\ref{fig:sample_profiles}.  Satellite galaxies are included in the
  model but are not shown for clarity. $\Sigma=7.0 \, (6.0, 5.0) \,
  \mathrm{M_{\odot} \, kpc^{-2}}$ corresponds to a V band surface brightness of
  $\sim24.8 \, (27.3, 29.8) \, \mathrm{mag\, arcsec^{-2}}$ assuming a
  mass-to-light ratio of 2.5 (see Fig.~\ref{fig:sample_profiles} for this
  approximate conversion scale). We have imposed a minimum surface density of
  $\Sigma = 3.0 \, \mathrm{M_{\odot} \, kpc^{-2}}$ ($\sim35 \, \mathrm{mag\,
  arcsec^{-2}}$) with Poisson noise to create the `background' in these images.
  The lumpy appearance of the diffuse light is due to shot noise in the
  adaptively smoothed particle distribution.}

  \label{fig:gallery}
\end{figure*}

\begin{figure*}
  \includegraphics[width=168mm, trim=0cm 0cm 0cm 0.0cm, clip=True]{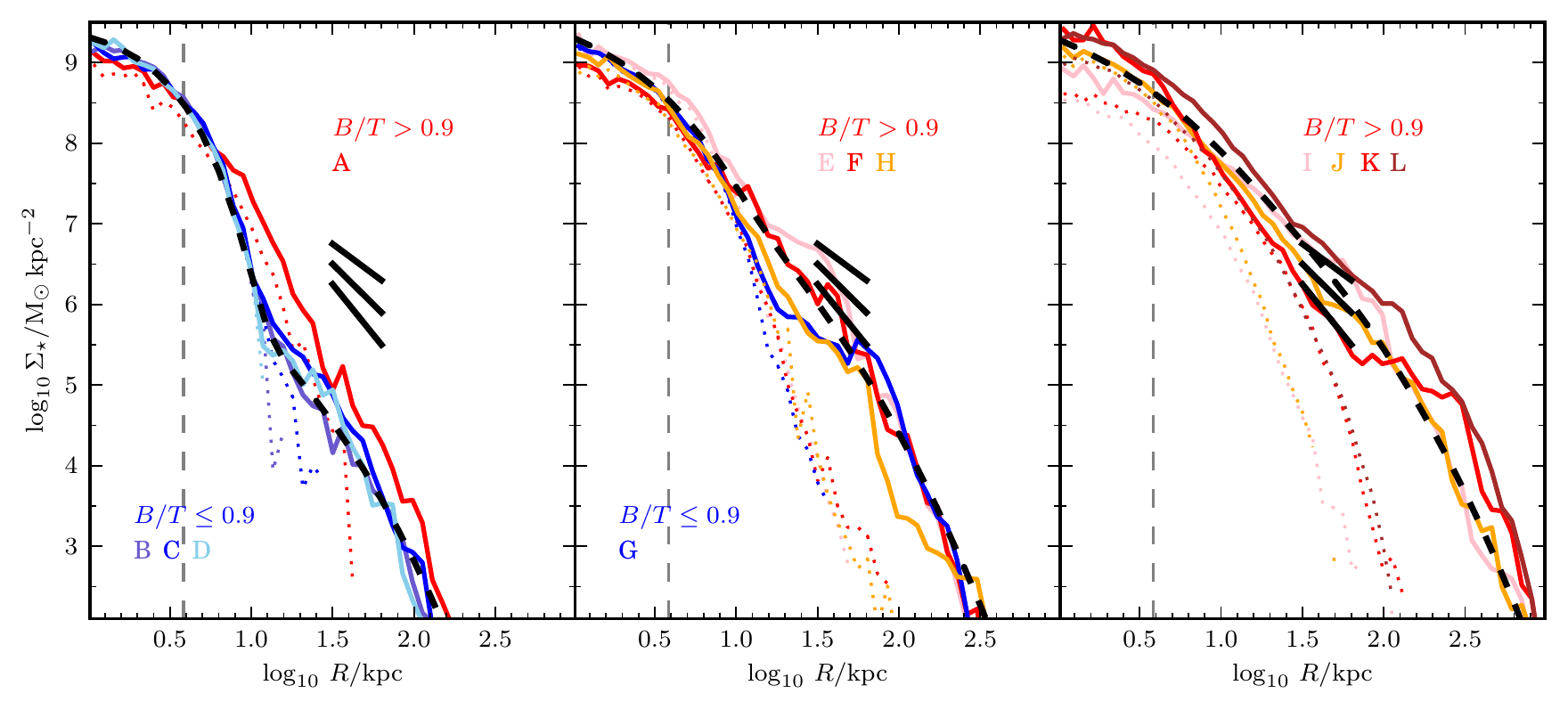}

  \caption{Solid lines show stellar mass surface density profiles for the
  galaxies shown in Fig.~\ref{fig:gallery} (identified by letters A--L). Line
  colours correspond to G11 B/T values as shown in the legend. Short black solid
  lines show slopes of -1.5, -2.0 and -2.5 for reference. The thick dashed black
  lines are the average total density profiles for galaxies with the same halo
  mass as the examples, discussed in section~\ref{sec:average_profiles}.  Dotted
  lines show the density of in situ stars only. A vertical dashed line marks the
  effective force softening scale.} 

  \label{fig:sample_profiles} 
\end{figure*}

\section{Results for individual galaxies}
\label{sec:individual}

Before we present surface density profiles of galaxies averaged in bins of
stellar mass and halo mass, we illustrate the basic output of our model with 12
examples of individual galaxies from our sample of 1872.

\subsection{Images}

Fig.~\ref{fig:gallery} shows 2D images of projected stellar mass surface
density for twelve of our simulated galaxies ($f_{\mathrm{mb}}=1\%$).  The
stellar mass associated with each tagged dark matter particle has been smoothed
over a cubic spline kernel with a scale radius enclosing its 64 nearest dark
matter neighbours \nocite{Springel05}(e.g. {Springel} 2005).  Satellite
galaxies (tagged particles in self-bound subhalos) have been removed, so the
remaining `lumps' on small scales are due to shot noise in the particle
distribution. 

The galaxies labelled A--D in Fig.~\ref{fig:gallery} represent the lowest halo
masses in our sample, with $M_{200} \approx 10^{12} \, \mathrm{M_{\odot}}$ and
$M_{\star} \approx 6\times10^{10} \, \mathrm{M_{\odot}}$ \nocite{Li08mwmass,
McMillan11}(similar to the Milky Way, e.g. Li \& White 2008; {McMillan} 2011).
Those labelled E--H are more massive isolated haloes with $M_{200} \approx
10^{12.5}\, \mathrm{M_{\odot}}$ and slightly larger $M_{\star}$, while those
labelled I--L represent the central galaxies of groups and poor clusters, with
$ 10^{13.25} < M_{200} < 10^{14} \, \mathrm{M_{\odot}}$ and $M_{\star}$ up to
$2\times10^{11} \, \mathrm{M_{\odot}}$.  The G11 model predicts that B, C, D, G
and K have $B/T<0.2$, and J has $B/T \sim 0.4$. The other examples fall into
our `early type' category defined by $B/T > 0.9$. 
 
The colour scale for surface mass density illustrates the observability of
different features. Regions with red colours are readily observable in the
Sloan Digital Sky Survey; $\mu_{r} < 25 \, \mathrm{mag \, arcsec^{-2}}$)
whereas those with yellow/green colours require deep imaging ($25 < \mu_{V} <
28 \, \mathrm{mag \, arcsec^{-2}}$, e.g.
\nocite{MD10}{Mart{\'{\i}}nez-Delgado}  {et~al.} 2010b). Careful reduction of
SDSS  images (in particular those in Stripe 82) can achieve $26 < \mu_{r} < 28
\, \mathrm{mag \, arcsec^{-2}}$ \nocite{Kaviraj10,Bakos12_arxiv}(e.g. {Kaviraj}
2010; {Bakos} \& {Trujillo} 2012) and the next generation of imaging surveys
will reach these depths routinely \nocite{lsstbook09}({LSST Science
Collaborations} {et~al.} 2009). Regions with $\mu_{r}\sim31 \, \mathrm{mag \,
arcsec^{-2}}$ (blue colours) are currently only accessible with resolved
starcounts in nearby galaxies
\nocite{BlandHawthorn05,Barker09,Bailin11,RadburnSmith11}(e.g.
{Bland-Hawthorn} {et~al.} 2005; {Barker} {et~al.} 2009; {Bailin} {et~al.} 2011;
{Radburn-Smith} {et~al.} 2011) and through stacking enembles of similar
galaxies \nocite{Zibetti04,Zibetti05,Tal11}({Zibetti} {et~al.} 2004, 2005;
{Tal} \& {van Dokkum} 2011).

\subsection{Individual density profiles}
\label{sec:sample_profiles}

The solid lines in Fig.~\ref{fig:sample_profiles} show the stellar mass surface
density profiles of the 12 galaxies from Fig.~\ref{fig:gallery}. Galaxies in
the three panels (representing different ranges of $M_{200}$ and $M_{\star}$)
show clear differences in the shape of their profiles, as well as the expected
differences in amplitude between galaxies of different mass.  Whereas there are
no clear differences in the appearance of the galaxies in the first two columns
of Fig.~\ref{fig:gallery}, the late type galaxies (B, C, D and G) clearly show
two structural components in Fig.~\ref{fig:sample_profiles}: an
exponentially-declining inner profile that breaks to a shallower slope at radii
between 10~kpc and 30~kpc. This break is not seen in the profiles of their
early type counterparts (A, E, F, H), which have roughly constant slope from
5~kpc to 100~kpc\footnote{The shape of the profile and the classification of
the galaxy as early or late type in the G11 model are not directly connected;
they are only correlated by the dependence of both on the mass accretion
history.}. 

The inner regions of these galaxies ($<10$~kpc) are very similar to each other,
even for different halo masses and at radii well outside the effective force
softening scale (shown by a vertical dashed line\footnote{Forces are
exactly Newtonian for particle separations greater than $2.8\epsilon = 3.84
\,\mathrm{kpc}$ with $h = 0.73$ and Plummer-equivalent softening length
$\epsilon = 1 \, h^{ - 1} \mathrm{kpc}$ fixed in comoving coordinates.}).
Coloured dotted lines in Fig.~\ref{fig:sample_profiles} show that in
situ stars dominate these regions.  Although the in situ component grows in
size with halo mass, the mass of the accreted component that dominates at
larger radii increases even more rapidly.  The surface density at
$\sim100$~kpc, which is almost entirely contributed by accreted stars in all
our examples, increases by three orders of magnitude across the panels. 

Therefore, the physical origin of the break mentioned above is the transition
from regions dominated by in situ stars to regions dominated by accreted stars.
This implies a clear connection between the typical shapes of galaxy
surface brightness profiles and the relative fraction and distribution of
accreted stars. The rest of this paper will focus on the importance of
accreted stars in changing the surface brightness profile shape as a function
of halo mass, using all the galaxies in our sample.

\begin{figure*}
  \includegraphics[width=164mm, clip=True]{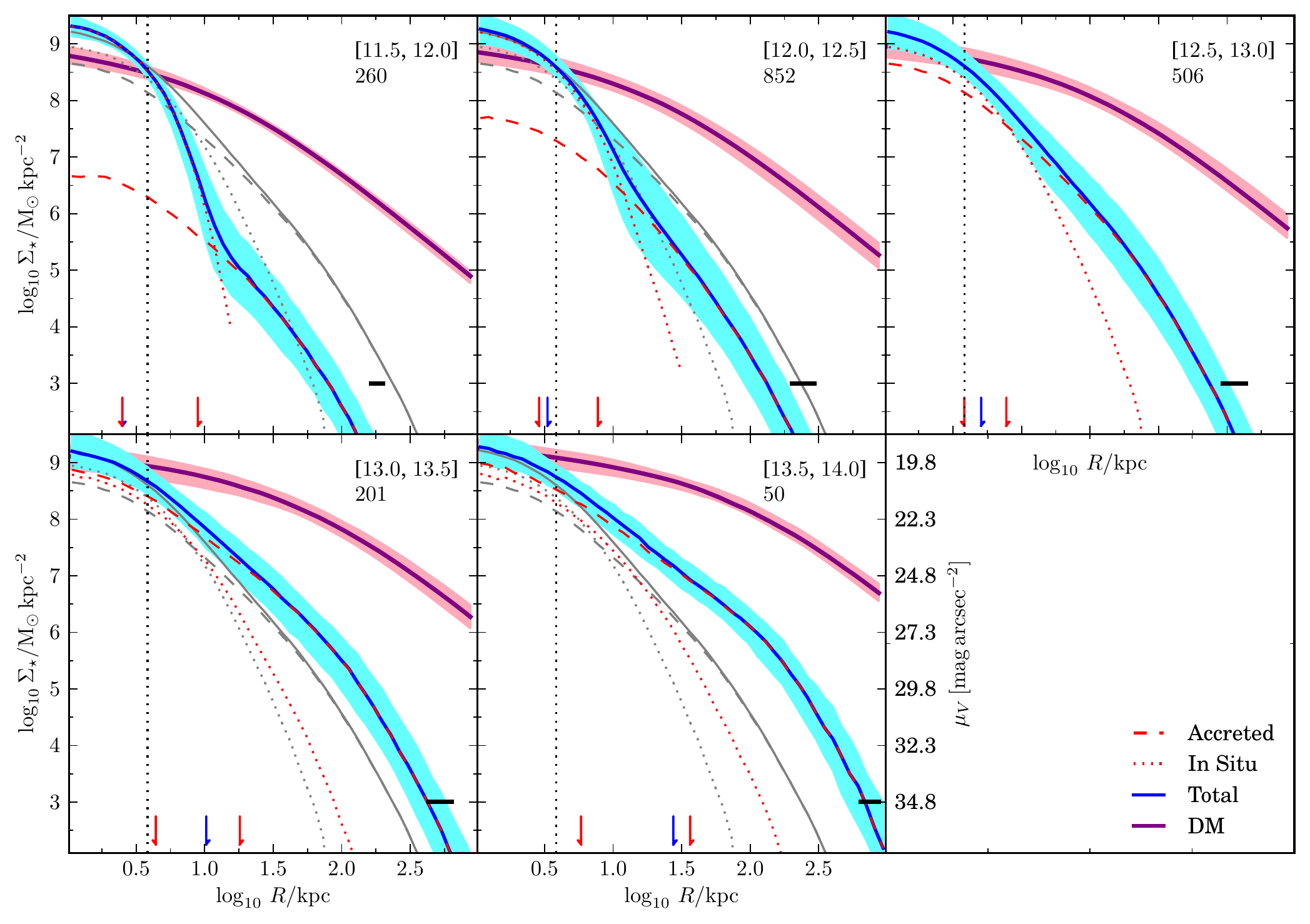} 

  \caption{Median profiles of circularly averaged stellar mass surface density,
  $\Sigma_{\star}$, for accreted stars (red dashed lines) and in situ stars
  (red dotted lines), in logarithmic bins of dark halo virial mass (range of
  $\log_{10}\,M_{200}/\mathrm{M_{\odot}}$ and number of galaxies per bin are
  shown in the top right of each panel). A blue solid line shows the median
  profile for $f_{\mathrm{mb}}=1\%$ combining accreted and in situ components;
  a light blue region indicates the 10--90 per cent scatter of the median
  profile. Arrows indicate half-mass radii of the median profiles (from left to
  right, in situ stars, all stars and accreted stars). Grey lines (dotted,
  dashed and solid) reproduce the corresponding red and blue lines from the
  $12.5<\log_{10}\,M_{200}/\mathrm{M_{\odot}}<13.0$ panel. A purple line and
  pink shading show the median dark matter density profile and its 10--90 per
  cent range.  A black horizontal bar shows the range of $R_{200}$ in each mass
  bin, and a vertical dotted black line indicates the effective softening scale
  $2.8\epsilon$. The scale on the right of the lower central panel gives an
  approximate conversion from $\Sigma_{\star}$ to surface brightness (in Vega
  magnitudes per square arcsecond) for the Johnson-Cousins V band, assuming
  $\Upsilon_{V} = M_{\star}/L_{V}=2.5$.} 
  \label{fig:density_profiles}
\end{figure*}

\begin{figure*} 
  
  \includegraphics[width=164mm, trim=0.0cm 0cm 0cm 0.0cm, clip=True]{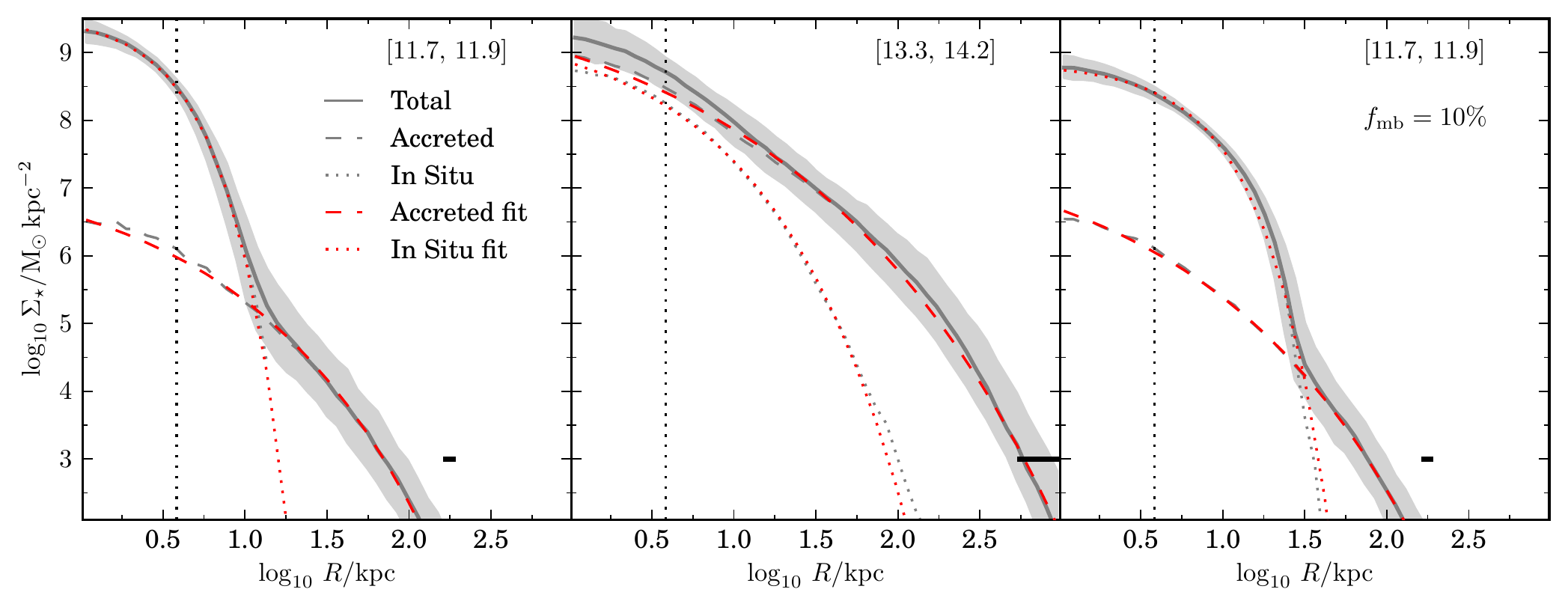}

  \caption{Average stellar mass surface density profiles as in
  Fig.~\ref{fig:density_profiles} for our 100 least massive haloes (left) and
  100 most massive haloes (centre) with $f_\mathrm{mb}=1\%$, and our 100 least
  massive haloes with $f_\mathrm{mb}=10\%$ (right). Legends indicate the
  corresponding range of $\log_{10} M_{200}/M_{\odot}$. Grey lines show our
  simulation results, red lines show \Sersic{} model fits to the accreted
  (dashed) and in situ (dotted) components, which overplot the simulation data
  almost everywhere. From left to right, the \Sersic{} parameters of the
  accreted star fits are $[\log_{10} \Sigma_{50}/\mathrm{M_{\odot} \,
  kpc^{-2}}, R_{50}/\mathrm{kpc}, n] = [5.31, 10.4, 2.56]$, $[7.21, 24.6,
  3.64]$ and $[5.37, 10.1, 2.96]$. The in situ star fits are $[9.00, 2.3,
  0.79]$, $[7.96, 5.5, 1.90]$ and $[8.21, 5.4, 0.88]$.}

  \label{fig:density_sersic_compare} 
\end{figure*}

\section{Average surface density profiles}
\label{sec:average_profiles}

Figure \ref{fig:density_profiles} shows the main results of this paper: the
median profiles of stellar mass surface density in circular annuli, for all the
galaxies in our sample in logarithmic bins of 0.5~dex in dark halo virial mass
($M_{200}$). In this figure we only show results for $f_{\mathrm{mb}}=1\%$,
for clarity. We do not take projections along the principal
axes of the galaxies or align them in any other way, so the relative
orientations of the profiles we combine are random.

The profiles of total stellar mass surface density (blue lines) show a tight
correlation in shape and amplitude with $M_{200}$, with 80 per cent of
profiles differing by no more than $0.5$~dex from the median at all radii in
all halo mass bins (a large part of this scatter, shown by the light blue band,
is due to the $0.5$~dex width of our halo mass bins). The separate
contributions of accreted and in situ stars explain the shape of the profiles
and their variation with halo mass. In situ stars (red dotted lines) dominate
the high surface brightness regions in haloes up to $\sim
10^{13}\mathrm{M_{\odot}}$, while accreted stars (red dashed lines) dominate at
radii greater than 30~kpc and in more massive haloes. Regardless of
$M_{200}$, neither the two subcomponents nor the combined stellar profile
follow the NFW distribution of the dark matter (purple lines).  

\makebold{Fig.~\ref{fig:density_sersic_compare} shows examples of
\nocite{Sersic68}{S\`{e}rsic} (1968) functions (c.f. \nocite{Graham05} Graham
\& Driver 2005) fit to our median in situ and accreted surface density
profiles. We find that both components are well described by such fits across
the entire $M_{200}$ range of our sample and for $1\%<f_{\mathrm{mb}}<10\%$. By
construction, the total profile is best fit by the sum of these two functions
-- a single \Sersic{} function is only an appropriate model for the {\em total}
surface density in haloes more massive than $M_{200}\sim10^{13}\, M_{\odot}$,
where the accreted component dominates at all radii.  Even in these haloes, the
in situ component makes a significant contribution, and a slight change in
shape due to the transition between accreted and in situ stars is still
apparent at $R<10$~kpc.} 

\begin{figure*}

  \includegraphics[width=164mm, clip=True, trim=0cm 0cm 0cm 0cm ]{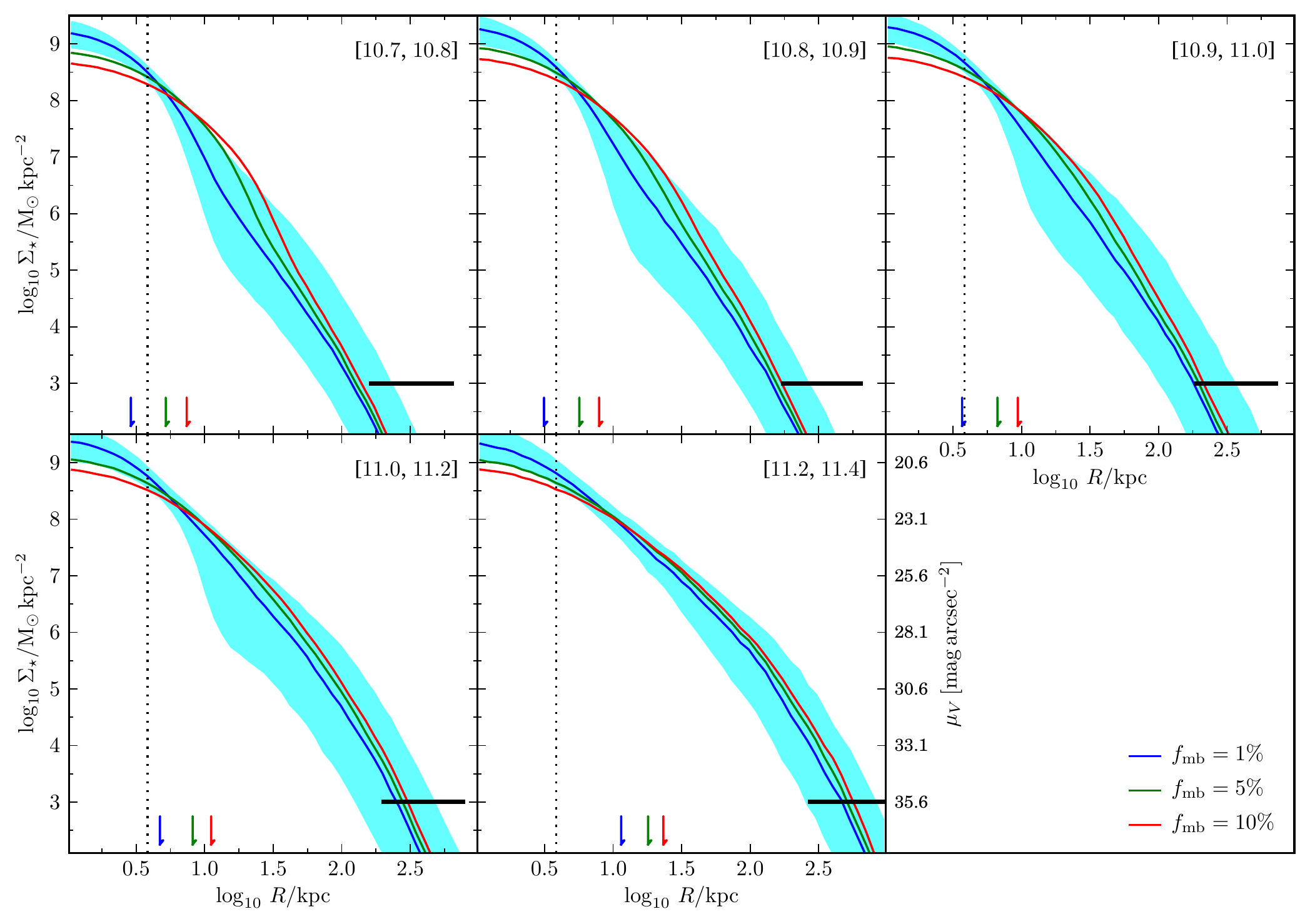}

  \caption{Surface density profiles in bins of total stellar mass,
  $M_{\star}$.  Note that the final two bins are wider than the
  first three. Different line colours correspond to different
  choices of $f_{\mathrm{mb}}$, which mainly affect the in situ
  component; the effect on the accreted component is negligible
  except in the most massive galaxies (see Appendix A).  The light
  blue region indicates the 10--90 per cent scatter of the median
  profile for the $f_{\mathrm{mb}}=1\%$ case.  Arrows show half
  mass radii and black lines show the range of $R_{200}$ in each
  bin.}

  \label{fig:density_profiles_mstar}
\end{figure*}

\begin{figure} 
  
  \includegraphics[width=79mm, trim=0.25cm 0.225cm 0.2cm 0.24cm, clip=True]{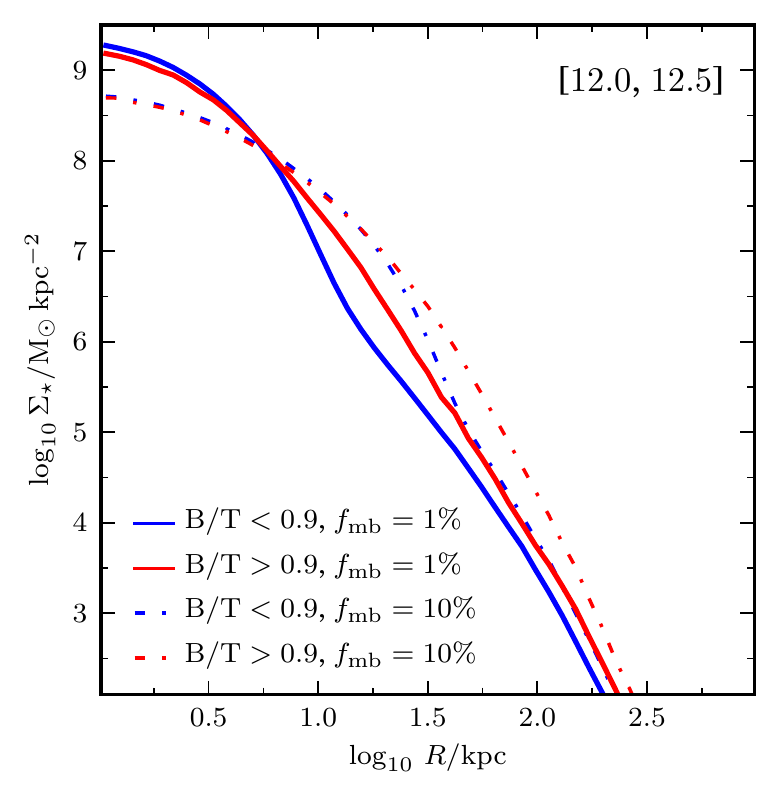}

  \caption{Average stellar mass surface density profiles for
  $f_{\mathrm{mb}}=1\%$ (solid lines) and $10\%$ (dot-dashed lines) in the halo
  mass bin $12.0<\log_{10} M_{200} <12.5$, subdivided into galaxies with $B/T
  <0.9$ ($N=704$, blue lines) and with $B/T > 0.9$ ($N=132$, red lines).}

  \label{fig:density_profiles_by_type} 
\end{figure}

Fig.~\ref{fig:density_profiles_mstar} repeats Fig.~\ref{fig:density_profiles}
but bins galaxies by stellar mass, $M_{\star}$, rather than $M_{200}$. The
strong trends with $M_{200}$ are less clear in the case of $M_{\star}$, except
in the most massive stellar mass bin. The cause of this is the considerable
scatter in the $M_{\star}$--$M_{200}$ relation \nocite{Guo10}(e.g. {Guo}
{et~al.} 2010).  Below their half-mass radii, the median surface brightness
profiles are hard to distinguish across a range of $10.7 <\log_{10} \,
M_{\star}< 11.2$.  A stronger variation of the profiles with $M_{\star}$ can be
seen at larger radii, $\sim30$--$100$~kpc.

Fig.~\ref{fig:density_profiles_mstar} also illustrates the effects of varying
$f_{\mathrm{mb}}$, over the range 1--10 per cent.  Relative to the 1 per cent
profiles (blue lines), the 10 per cent profiles (red lines) are $\sim30$ per
cent more extended and have lower central surface density (see
section~\ref{sec:fmb}). This rescaling of the in situ component is the main
effect of changing $f_{\mathrm{mb}}$. The most significant impact of higher
$f_{\mathrm{mb}}$  on the accreted component is in the very outer regions of
the most massive galaxies, where slightly more stars are found on weakly bound
orbits (perhaps as the result of earlier stripping). Again we conclude that our
density profile results do not depend strongly on the value of
$f_{\mathrm{mb}}$, at least within a range consistent with the observed
mass--size relation. 

\begin{figure*} 

  \includegraphics[width=180mm, clip=True, trim=0cm 0cm 0cm 0cm ]{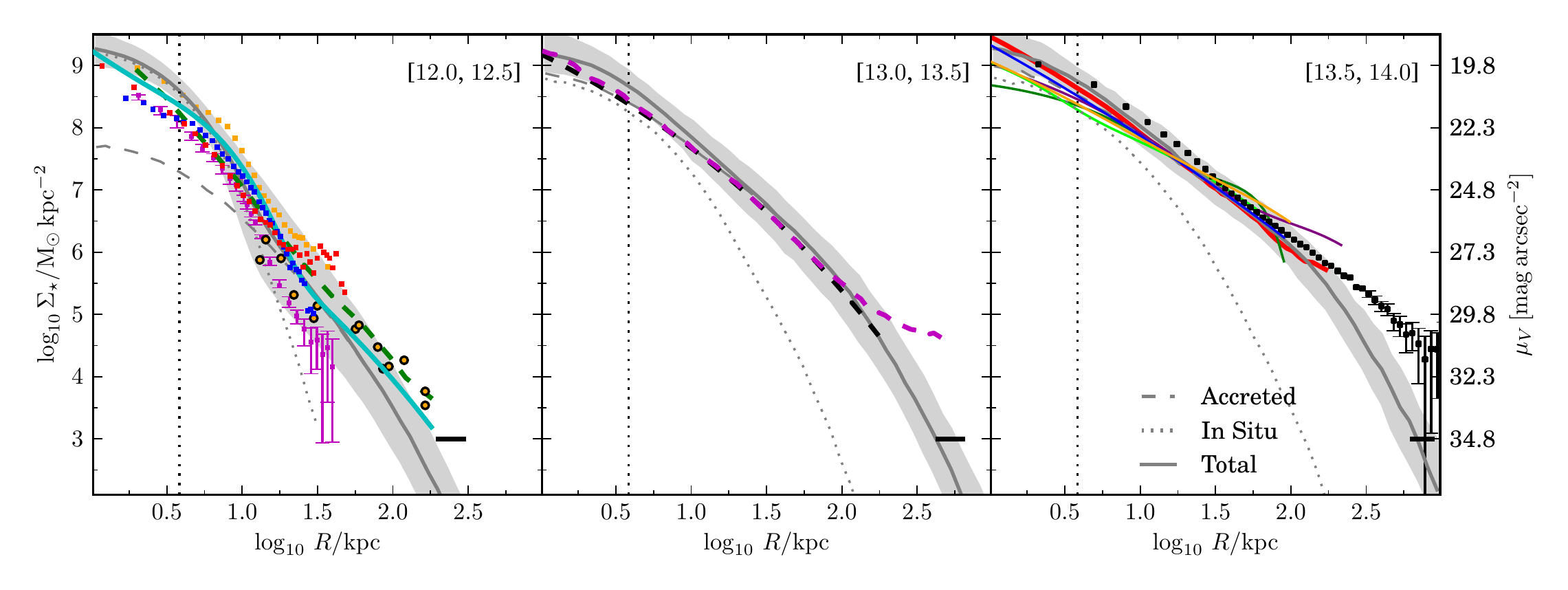}

  \caption{As Fig.~\ref{fig:density_profiles}.  The surface density profiles in
  each bin of halo mass for $f_{\mathrm{mb}}=1\%$ only are shown in grey, and
  compared with observational data (and other simulations) given in
  Table~\ref{tab:obs}. The assignment of galaxies to halo mass bins for
  the observational data is approximate.}

  \label{fig:density_profiles_obs}
\end{figure*}

\begin{table*}

  \caption{Surface mass density profile data shown in Fig.~\ref{fig:density_profiles_obs}. From left to
  right, columns give: the range of halo mass to which the data (or simulations)
  are compared ($\log_{10} M_{200}/M_{\odot}$, corresponding to panels in
  Fig.~\ref{fig:density_profiles_obs}); the target galaxy or galaxies; the
  source of the data; the symbol or line style used in
  Fig.~\ref{fig:density_profiles_obs}; the photometric bandpass of the data;
  the stellar mass-to-light ratio we have assumed (where the original authors
  do not present their results in terms of stellar mass surface density); and
  comments on the data (LTG: late type galaxy).}

  \label{tab:obs}
  \begin{tabular}{@{}lllllll}
    \hline

    Halo mass & Galaxies & Reference & Marker & Band & $M_{\star}/L$ & Comments \\ 
    \hline
    $[12.0, 12.5]$ & M81       & \nocite{Barker09}{Barker} {et~al.} (2009)      & Orange squares & $V$     & 2.5 & LTG \\
    & M31       & \nocite{Gilbert09}{Gilbert} {et~al.} (2009)     & Orange circles & $V$     & 2.5 & LTG \\
    & M31       & \nocite{Courteau11}{Courteau} {et~al.} (2011)    & Cyan line      & $I$     & 1.5 & LTG, composite profile from various sources \\
    & NGC 1087  & \nocite{Bakos12_arxiv}{Bakos} \& {Trujillo} (2012) & Blue squares   & $ugriz$ & --  & LTG, $\Sigma_{\star}$ from authors \\
    & NGC 7716  & \nocite{Bakos12_arxiv}{Bakos} \& {Trujillo} (2012) & Red squares    & $ugriz$ & --  & LTG, $\Sigma_{\star}$ from authors \\
    & NGC 2403  & \nocite{Barker12}{Barker} {et~al.} (2012)      & Magenta squares & $V$     & 2.5 & LTG \\
    & \textit{GIMIC} & \nocite{Font11}{Font} {et~al.} (2011) & Green dashes & $V$     & 2.5 & Stack of $\sim 400$ simulated LTGs \\ 
    \hline
    $[13.0, 13.5]$ & LRG stack    & \nocite{Tal11}{Tal} \& {van Dokkum} (2011)            & Magenta line   & $r$     & 2.0 & Stack, $N=42579$, $\langle z \rangle \sim 0.34$ \\
    & OBEY sample  & \nocite{Van-Dokkum10}{van Dokkum} {et~al.} (2010)     & Black line     & $r$     & 2.0 & \Sersic{} profile fit to stacked \nocite{Tal09}{Tal} {et~al.} (2009) data \\
    \hline
    $[13.5, 14.0]$ & NGC 6173  & \nocite{Seigar07}{Seigar}, {Graham} \& {Jerjen} (2007)   & Blue line        & $R$     & 2.0 & BCG, Abell 2197    \\
    & UGC 9799  & \nocite{Seigar07}{Seigar} {et~al.} (2007)   & Dark green line  & $R$     & 2.0 & BCG, Abell 2052    \\
    & NGC 3551  & \nocite{Seigar07}{Seigar} {et~al.} (2007)   & Orange line      & $R$     & 2.0 & BCG, Abell 1177    \\
    & GIN 478   & \nocite{Seigar07}{Seigar} {et~al.} (2007)   & Purple line      & $R$     & 2.0 & BCG, Abell 2148    \\
    & NGC 4874  & \nocite{Seigar07}{Seigar} {et~al.} (2007)   & Light green line & $R$     & 2.0 & BCG, Coma cluster  \\
    & M87       & \nocite{Kormendy09}{Kormendy} {et~al.} (2009) & Red line         & $V$     & 2.5 & BCG, Virgo cluster \\
    & BCG stack & \nocite{Zibetti05}{Zibetti} {et~al.} (2005)  & Black squares    & $i$     & 1.5 & Stack of SDSS MaxBCG clusters, richness $>15$ 
  \end{tabular}
\end{table*}

In Section~\ref{sec:sample_profiles} we discussed an interesting relationship
between the bulge-to-total stellar mass ratio $B/T$ (predicted by the G11
model) and the shape of the surface density profiles we obtain with particle
tagging. In Figure~\ref{fig:density_profiles_by_type} we show that this result
holds on average, plotting the $f_{\mathrm{mb}}=1\%$ profile in the bin
$12.0<\log_{10} M_{200}<12.5$ (top-centre panel of Figure
\ref{fig:density_profiles}) separately for galaxies with $B/T >0.9$ and
$B/T<0.9$. The latter profile shows a clear inflection at $\sim10$~kpc,
which corresponds to the radius at which accreted stars begin to
dominate over in situ stars. The absence of this break in galaxies with $B/T
>0.9$ reflects a greater contribution of accreted stars to the inner
parts of galaxies and the action of violent relaxation in major mergers (the
primary cause of high $B/T$ in the G11 model), which makes the profiles
of the two components more similar.  Figure~\ref{fig:density_profiles_by_type}
shows that profiles of galaxies with low and high G11 $B/T$ can still
be distinguished if we use a much higher value of $f_{\mathrm{mb}}$, although
the accreted-in situ break is then much less clear. We caution that the
value of $B/T$ is calculated in a simple manner and is influenced by other
processes in the G11 model (such as disc instabilities and starbursts). In
individual cases it does not correspond perfectly to a separation between
two-component and quasi-power-law profiles in our tagged particle model.

\section{Comparison with literature data}
\label{sec:obs}

\begin{figure*}

  \includegraphics[width=168mm, clip=True, trim=0cm 0cm 0cm 0cm ]{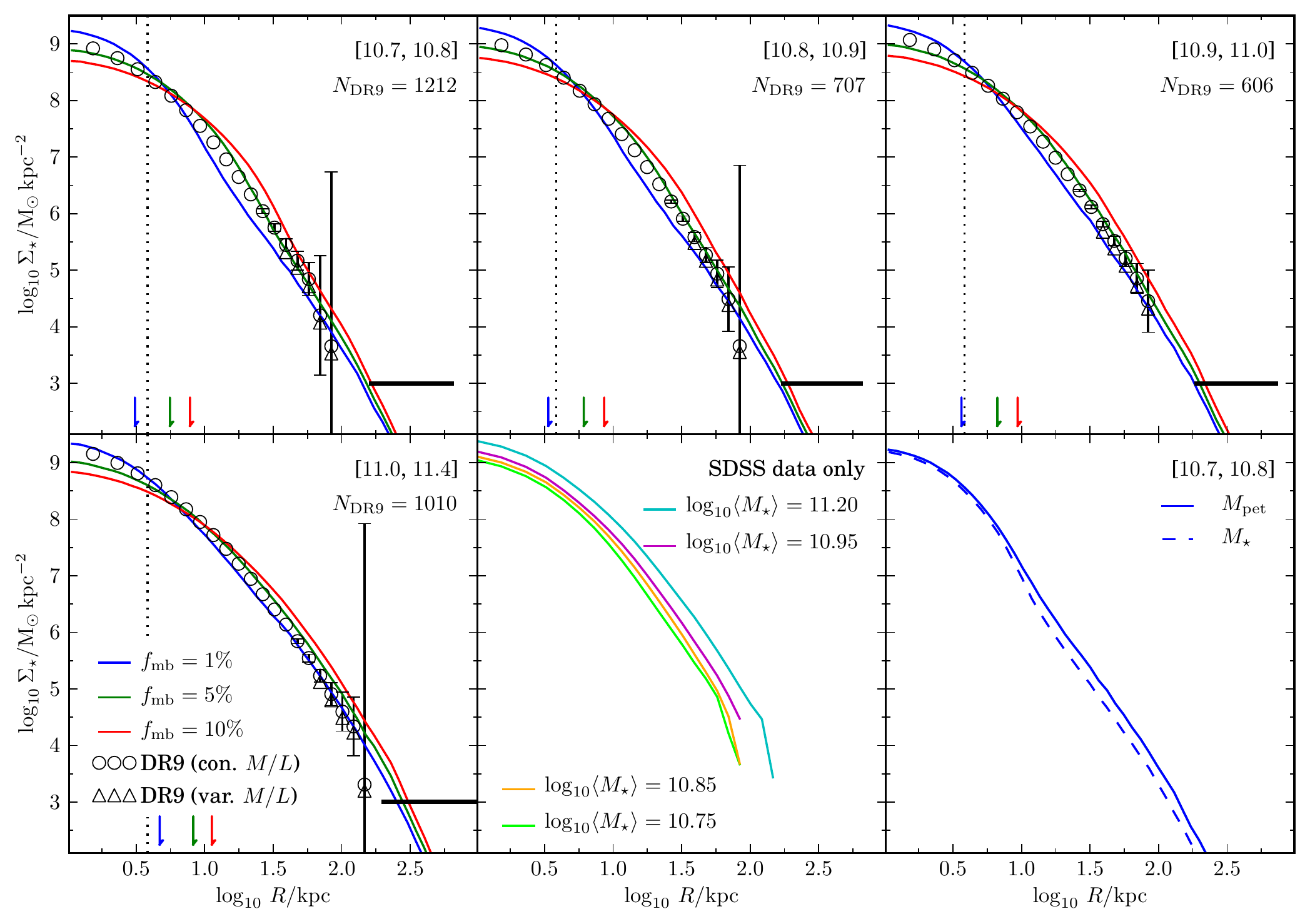}

  \caption{Symbols show the average stellar mass surface density profiles
  obtained from stacks of SDSS DR9 $r$ band images as described in the text and
  appendix~\ref{appendix_d}, assuming a constant stellar mass to light ratio
  (open circles) and, where significantly different, a colour-dependent $M/L$
  (open triangles). Error bars approximate `$1\sigma$' of the distribution of
  uncertainty in the average profiles combining Poisson errors in flux
  measurement with the sample variance of the stack ($N_{DR9}$ given in each
  panel shows the number of galaxies in the bin). Coloured lines (blue, green
  and red) show stacks made from our simulations as in
  Fig.~\ref{fig:density_profiles_mstar} but here binning galaxies by their
  Petrosian mass $M_{\mathrm{pet}}$ (see text). The lower central panel shows
  the four SDSS profiles only (colours indicate the central mass of each bin).
  \makebold{The lower right panel reproduces the $f_{\mathrm{mb}}=1\%$ profile
  from the $[10.7,10.8]$ panel (solid line) and compares it to the average
  profile of galaxies stacked in the same range of total stellar mass
  $M_{\star}$ (dashed line) rather than $M_{\mathrm{pet}}$.}} 
  
  \label{fig:density_profiles_sdss}
\end{figure*}

In Fig.~\ref{fig:density_profiles_obs} we compare the surface density profiles
shown in Fig.~\ref{fig:density_profiles} (binned by $M_{200}$) with deep
observational data from a variety of sources (summarised in
Table~\ref{tab:obs}). We do this to illustrate the variety of different surface
density profiles that have been reported in the literature, rather than to
match any particular observation. In most cases our choice of an $M_{200}$ bin
for each observational dataset is based on the observed stellar mass
\nocite{Guo10, Moster10}(e.g. {Guo} {et~al.} 2010; {Moster} {et~al.} 2010) and
is thus very rough.  Where authors have presented their data in terms of
stellar mass surface density, we use their values directly.  Otherwise,
since we find that all but our lowest-mass simulated galaxies have very
shallow $M_{\star}/L$ gradients at $R>10$~kpc, we assume a galaxy-wide
$M_{\star}/L$ appropriate to each bandpass (listed in Table~\ref{tab:obs},
based on the ages and metallicities of our simulated stars and the models of
\nocite{BC03}{Bruzual} \& {Charlot} 2003 with a \nocite{Chabrier03}{Chabrier}
2003 IMF). There are many systematic differences between these datasets
(including photometric bandpass, surface brightness dimming corrections,
$K$-corrections, cosmology and in some cases, the choice of IMF).  We have
attempted to correct for these differences where necessary.  Such corrections
amount to less than $0.1$~dex in most cases.

In the halo mass range $12.0 < \log_{10} M_{200}/\mathrm{M_{\odot}} < 12.5$ we
show data from galaxies comparable to the Milky Way and M31
($M_{200}\sim10^{12} \mathrm{M_{\odot}}$, e.g.  \nocite{Watkins10}{Watkins}
{et~al.} 2010). The composite $I$ band profile of M31 from
\nocite{Courteau11}{Courteau} {et~al.} (2011, cyan line) agrees well with the
average profile of accreted stars in our model (dashed grey line) for $R
\lesssim 50$~kpc. At $R>100$~kpc  {Courteau} {et~al.} find a higher surface
density than our model; this portion of their profile is based on the
individual fields of \nocite{Gilbert09}{Gilbert} {et~al.} (2009; orange dots),
some of which may contain substructure.
The galaxies M81 \nocite{Barker09}({Barker} {et~al.} 2009), NGC 2403
($M_{\star}\sim10^{10}\mathrm{M_{\odot}}$; \nocite{Barker12}{Barker} {et~al.}
2012), NGC 1087 ($M_{\star}\sim10^{10.4}M_{\odot}$) and NGC 7716
($M_{\star}\sim10^{10.5}M_{\odot}$; \nocite{Bakos12_arxiv}{Bakos} \& {Trujillo}
2012) show similar profiles. All show hints of breaking to a shallower slope
beyond 15--20~kpc, although these upturns occur close to the limiting depth of
the observations. 

In the same panel we compare with the GIMIC SPH simulations
\nocite{Crain09}({Crain} {et~al.} 2009), the only other large cosmological
simulation of stellar haloes in this $M_{200}$ range (though see also
\nocite{Croft09}{Croft} {et~al.} 2009), with a particle mass $7.7\times$ larger
than Millennium~II and the same force softening length.  \nocite{Font11}{Font}
{et~al.} (2011) stacked $\sim400$ galaxies from GIMIC. They find circularly
averaged density profiles that are well described by a concentrated in situ
component and a diffuse accreted component. The transition between the two
components is less obvious in their profiles than in our $f_{\mathrm{mb}}=1\%$
model, and the shallower outer slope seen in the GIMIC simulations is in better
agreement with the M31 data of \nocite{Gilbert09}{Gilbert} {et~al.} (2009) at
$\sim100$~kpc.

In the range $13.0 < \log_{10} M/\mathrm{M_{\odot}} < 13.5$ accreted
stars dominate the overall surface density profile. We compare with the
stacked surface brightness profile of $\sim42000$ SDSS LRGs (luminous red
galaxies, thought to be mostly group/cluster centrals; $M_{\star}\sim10^{11}\,
\mathrm{M_{\odot}}$, $M_{200} \sim 10^{13.2} \, \mathrm{M_{\odot}}$) at
$\langle z \rangle \sim 0.34$ from \nocite{Tal11}{Tal} \& {van Dokkum} (2011).
These data are a good match to the surface density profiles of our simulated
galaxies from $10$--$100$~kpc.  Below $10$~kpc the simulated profile is steeper
than the data, although this region is sensitive to our treatment of the in
situ component.  The simulation does not reproduce the upturn in the observed
profile at $\sim100$~kpc, which \nocite{Tal11}{Tal} \& {van Dokkum} (2011)
attribute to limitations in their correction for residual light from unresolved
companion galaxies. Our model supports this interpretation because it does not
predict a separate physical component with a shallow density profile that could
explain the upturn. We also show the stacked profiles of 14 nearby early type
galaxies from the volume-limited OBEY survey \nocite{Tal09}({Tal} {et~al.}
2009). This stack includes all ellipticals in the survey with stellar mass
$\log_{10} M_{\star}/\mathrm{M_{\odot}}=11.45 \pm 0.15$ (appendix D of
\nocite{Van-Dokkum10}{van Dokkum} {et~al.} 2010). These data match the
\nocite{Tal11}{Tal} \& {van Dokkum} (2011) stack at small radii and do not show
an excess over our simulated profile at $R\gtrsim100$~kpc, possibly because
\nocite{Van-Dokkum10}{van Dokkum} {et~al.} (2010) stack 2D fits to the OBEY
galaxies rather than stacking the images directly.

Finally, the most massive haloes in our simulation have $13.5 < \log_{10}
M_{200}/\mathrm{M_{\odot}} < 14.0$. We compare these with individual BCG
profiles from \nocite{Kormendy09}{Kormendy} {et~al.} (2009, M87) and
\nocite{Seigar07}{Seigar} {et~al.} (2007, Abell cluster cD galaxies; we plot
their `double \Sersic{}' fits).  Our simulations match these profiles
well from $10$--$100$~kpc, although some of the observed galaxies could belong
to haloes more massive than the median of our bin (e.g.  $\log_{10} \,
M_{200}/\mathrm{M_{\odot}} = 15.1$ for NGC 4874/Coma and $14.3$ for UGC
9799/Abell 2052; \nocite{Reiprich02}{Reiprich} \& {B{\"o}hringer} 2002). We
also show the results of \nocite{Zibetti05}{Zibetti} {et~al.} (2005) who
stacked SDSS images for BCGs at $\langle z \rangle \sim0.25$. This is deeper
than any of the individual profiles and shows a clear excess over our models
beyond 100~kpc. Some of this disagreement may be due to a mismatch in
the average $M_{200}$ of the galaxies being compared, as mass--richness
relations suggest that the typical halo mass of the Zibetti et al. sample is
$M_{200} \gtrsim 10^{14.2}\, \mathrm{M_{\odot}}$ \nocite{Rozo09}(e.g. {Rozo}
{et~al.} 2009). It may also indicate that effects neglected by our model become
important in this regime.  For example, it may be that the angular momentum of
disc galaxies affects the orbital energy distribution of their tidal tails, or
that ram-pressure stripping of cold gas leads to more rapid disruption of
satellites; in both cases, more stripped stars would be deposited on weakly
bound orbits with large apocentres.

\section{Comparison with stacked SDSS data} \label{sec:sdss}

It is more appropriate to compare the average galaxy surface density profiles
from our models with similar averages constructed from large galaxy samples
(such as the stacking analyses of \nocite{Tal11}{Tal} \& {van Dokkum} 2011 and
\nocite{Zibetti05}{Zibetti} {et~al.} 2005) than with individual galaxies or small surveys
as we did in the previous section. We have carried out our own simple
stacking analysis of massive galaxies observed by the Sloan Digital Sky Survey
(SDSS) Data Release 9 \nocite{sdssdr9}(DR9; {Ahn} {et~al.} 2012) in bins of stellar mass (as given
by the MPA-JHU Value-Added
Catalogue\footnote{\url{http://www.mpa-garching.mpg.de/SDSS/DR7}}).  Our method
for constructing stacked images is described in appendix~\ref{appendix_d}.

\begin{figure*} 

  \includegraphics[width=148mm, clip=True]{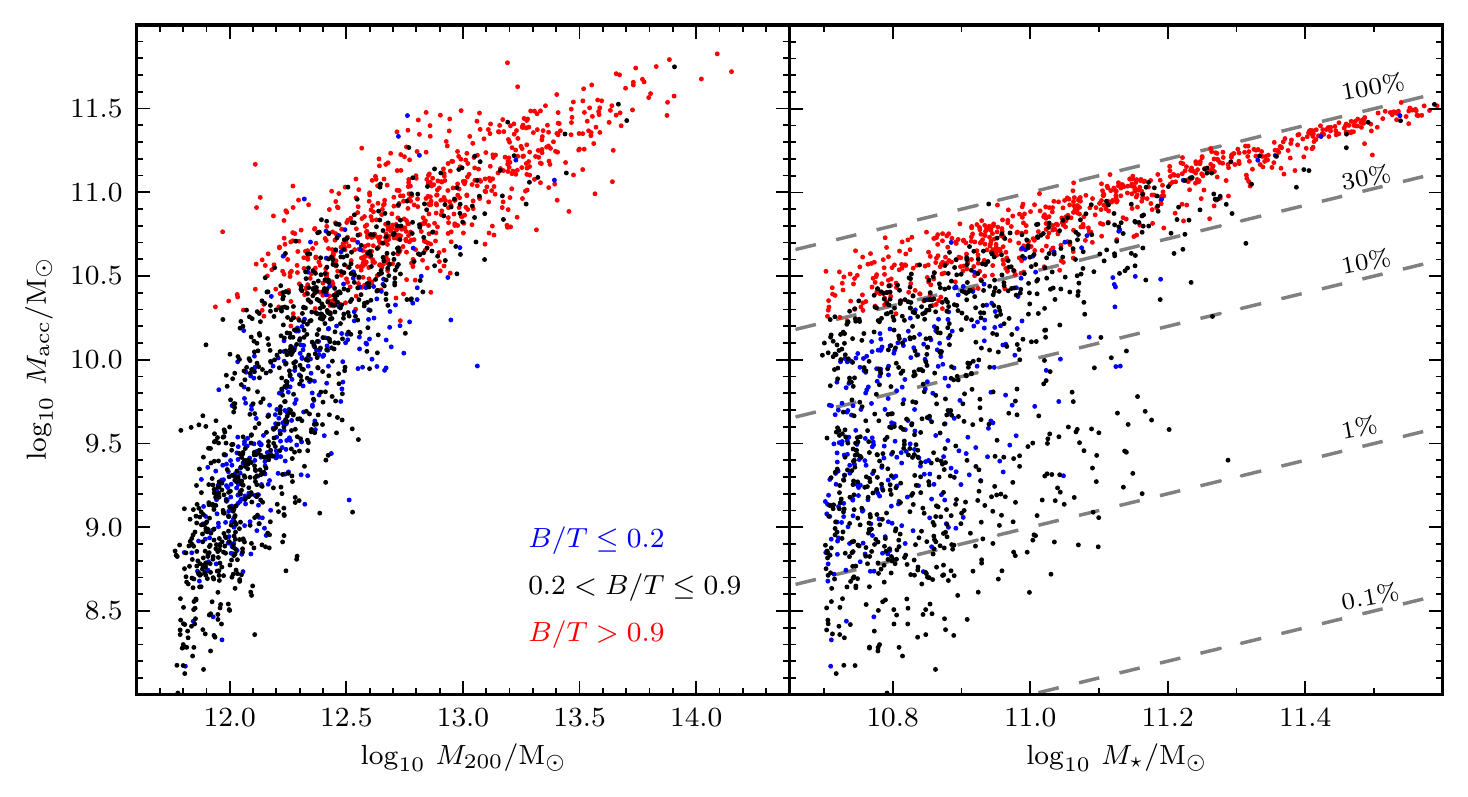}

  \caption{Left: points show the total mass of accreted stars,
  $M_{\mathrm{acc}}$ as a function of $M_{200}$, split by bulge-to-total ratio
  as shown in the legend. Right: points show $M_{\mathrm{acc}}$ as a function
  of stellar mass, $M_{\star}$.  Grey lines correspond to fractions of the
  total stellar mass as indicated.} 

  \label{fig:acc_mass} 
\end{figure*}

The resulting density profiles are shown as open circles in
Fig.~\ref{fig:density_profiles_sdss}, split into four bins of stellar mass,
each of which is obtained from a stack of  $N_{DR9}$ galaxies as indicated. The
panel labelled `SDSS data only' summarises these four profiles, showing a clear
shift in amplitude from the least to the most massive bin, out to the largest
measured radius. Each panel assumes a constant $r$ band mass-to-light ratio
(the average of the MPA-JHU $M/L_{r}$ values in the corresponding MPA-JHU mass
bin, ranging from $2.3$ to $2.8$ from the first to last bin; see
appendix~\ref{appendix_d}) but this result holds even if the same $M/L_{r}$ is
used for all panels, or if a colour-dependent $M/L_{r}$ relation is used
\nocite{Bell03}({Bell} {et~al.} 2003). However, the \makebold{stacked
observational data} do not show any significant change in profile shape of the
kind seen in previous figures (including observations of individual galaxies). 

Each panel compares our SDSS stacks to the average profiles of simulated
galaxies\footnote{The observed and simulated profiles in
Fig.~\ref{fig:density_profiles_sdss} should only be compared at
$R\gtrsim5$~kpc. At smaller radii, the point spread function, which we have not
deconvolved, dominates the observed profiles and numerical softening affects
the simulated profiles.} binned by their Petrosian stellar mass,
$M_{\mathrm{pet}}$ (blue, green and red lines for $f_{\mathrm{mb}}=1\%, 5\%$
and $10\%$, respectively).  We use $M_{\mathrm{pet}}$ rather than the true
stellar mass $M_{\star}$ in order to reproduce approximately the bias
introduced by SDSS \texttt{modelMag} magnitudes, from which the MPA-JHU masses
are derived (see appendix~\ref{appendix_d}). \makebold{$M_{\mathrm{pet}}$ is
always an underestimate of $M_{\star}$, thus the galaxies included in each bin
of Fig.~\ref{fig:density_profiles_sdss} are different from those in the
corresponding bin of Fig.~\ref{fig:density_profiles_mstar}. The result is a
small systematic increase in the amplitude of the average density profile, by
up to $\sim 0.5$~dex at large radii. This effect is most evident in the $10.7 <
\log_{10} M_{\mathrm{pet}} < 10.8$ bin for $f_{\mathrm{mb}}=1\%$, where it
obscures the inflection seen in Fig.~\ref{fig:density_profiles_mstar} (see
lower right panel of Fig.~\ref{fig:density_profiles_sdss}).  The weakness of
the overall trend in Fig.~\ref{fig:density_profiles_sdss} compared to that in
Fig.~\ref{fig:density_profiles} is mainly due to scatter between $M_{\star}$
and $M_{200}$ (see Fig.~\ref{fig:density_profiles_mstar}).  For example, the
bin $10.7 < \log_{10} M_{\star} < 10.8$ corresponds to a $\sim1.5$ dex range in
$M_{200}$}.

Even though the trends in the simulated data are quite weak, for
$f_{\mathrm{mb}}=1\%$ they are still clearly stronger than observed.  For
$f_{\mathrm{mb}}=5\%$, on the other hand, the agreement with observation is
quite good, and a slightly smaller value of $f_{\mathrm{mb}}$ would agree even
better, consistent with our findings in Section~\ref{sec:size_mass}.

\section{Origin and structure of stellar haloes}
\label{sec:origin}

Having demonstrated that our particle-tagging model produces surface density
distributions that agree reasonably well with observations, we now use it to
examine the origin of stellar haloes as well as the relationships between the
properties of central galaxies and the structure of their diffuse light.

The left panel of Fig.~\ref{fig:acc_mass} shows $M_{\mathrm{acc}}$, the
total mass of accreted stars in each of our galaxies, as a function of virial
mass, $M_{200}$.  The right panel shows $M_{\mathrm{acc}}$ as a function of
the total stellar mass of the system, $M_{\star}$.  $M_{\mathrm{acc}}$
increases much less steeply with $M_{200}$ above a characteristic mass
$M_{200}\sim10^{12.5}\, \mathrm{M_{\odot}}$. This corresponds roughly to the
transition mass  predicted by the G11 model, above and below which the galaxy
mass function is dominated by early and late type galaxies respectively. We see
in Fig.~\ref{fig:acc_mass} that elliptical (G11 $B/T > 0.9$) and late type
galaxies have a clean separation at an accreted stellar mass of
$M_{\mathrm{acc}}\sim3\times10^{10}\, \mathrm{M_{\odot}}$, which (as shown in
the right panel of Fig.~\ref{fig:acc_mass}) corresponds to $\sim30$ per
cent of the total (accreted plus in situ) central galaxy stellar mass
$M_{\star}$ (for all $M_{200}$). This simply reflects the fact that the `major
merger' criterion for the destruction of discs (hence formation of elliptical
galaxies) in G11 is a progenitor mass ratio of 1:3 or lower.

\nocite{Purcell07}{Purcell} {et~al.} (2007) made similar predictions for
$M_{\mathrm{acc}}$ as a function of $M_{200}$ at $z=0$ using prescriptions for
halo assembly histories and subhalo orbital properties based on  numerical
simulations.  As their results demonstrated, the relationship between
$M_{\mathrm{acc}}$ and $M_{200}$ seen in Fig.~\ref{fig:acc_mass} is a natural
outcome of the CDM progenitor halo mass function and the relation between
$M_{\star}$--$M_{200}$, which is thought to be roughly monotonic
\nocite{WhiteFrenk91,Benson00,vandenBosch03}(e.g. {White} \& {Frenk} 1991;
{Benson} {et~al.} 2000; {van den Bosch}, {Yang} \&  {Mo} 2003). To first order,
this means that $M_{\mathrm{acc}}$ is set by the typical ratio of $M_{200}$
between the few most massive accreted progenitors and the main halo
($\sim0.1$--$1\%$).  However, $M_{\star}$--$M_{200}$ relations derived from
galaxy abundance matching (like those predicted by semi-analytic models) show
an inflection corresponding to a peak in galaxy formation efficiency at
$M_{\mathrm{peak}} \approx M_{200}\sim10^{12}\, \mathrm{M_{\odot}}$
\nocite{Eke04, Conroy06, Vale06, Guo10, Moster10}(e.g. {Eke} {et~al.} 2004;
Conroy, Wechsler \& Kravtsov 2006; Vale \& Ostriker 2006; {Guo} {et~al.} 2010;
{Moster} {et~al.} 2010). This creates two regimes in the scaling of
$M_{\mathrm{acc}}$ with $M_{200}$.

In haloes with $M_{200} \lesssim M_{\mathrm{peak}}$, in situ star formation
efficiency per unit halo mass increases steeply along the halo mass function,
up to a maximum around the Milky Way mass \nocite{Moster10}(e.g. {Moster}
{et~al.} 2010). Because even the most massive accreted haloes typically have
much lower galaxy formation efficiency than the main halo, $M_{\mathrm{acc}}$
remains a small fraction of $M_{\star}$ and depends strongly on
$M_{200}$\footnote{In the extreme case, sharp thresholds for galaxy formation
at very low $M_{200}$ (the atomic hydrogen cooling limit, reionization etc.)
will result in central galaxies that do not accrete any luminous progenitors
and have $M_{\mathrm{acc}}=0$.}. For $M_{200} \gtrsim 10 M_{\mathrm{peak}}$,
the most massive progenitor haloes have a galaxy formation efficiency
comparable to or even higher than the main halo and $M_{\mathrm{acc}}$ makes up
a much larger fraction of $M_{\star}$. However, as the $M_{\star}$--$M_{200}$
relation flattens, the number of progenitors continues to scale with $M_{200}$,
but the stellar mass per massive progenitor does not. The result is that
$M_{\mathrm{acc}}$ scales more slowly with $M_{200}$ than it does below
$M_{\mathrm{peak}}$. 

\begin{figure} 

  \includegraphics[width=84mm, trim=0.0cm 0.2cm 0.2cm 0.2cm, clip=True]{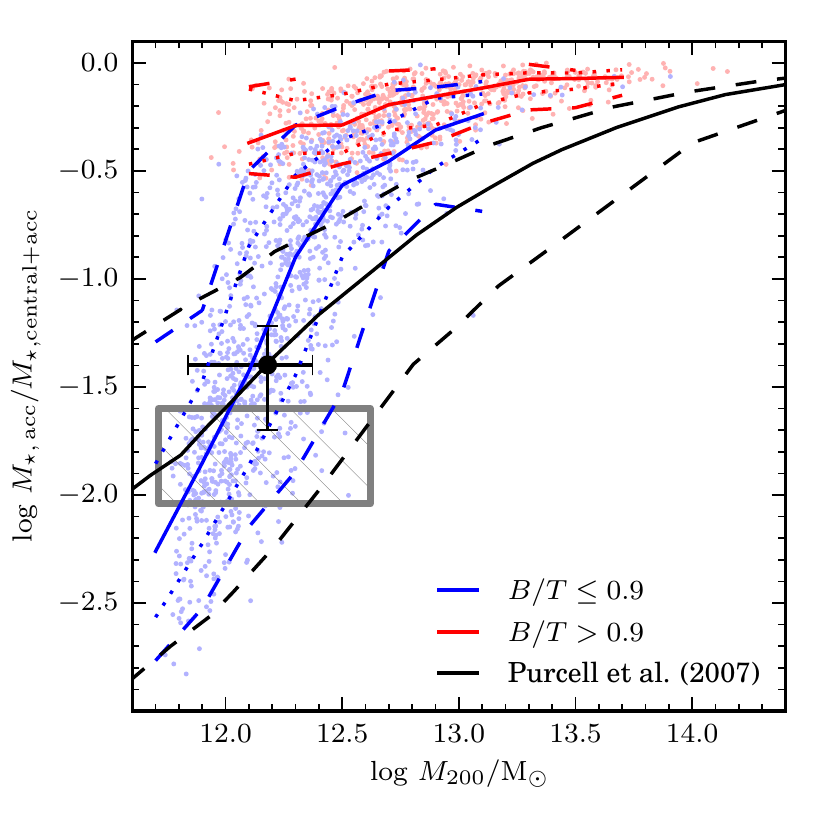}

  \caption{Solid blue and red lines show the median ratio of accreted stellar
  mass (both halo and bulge) to the total galaxy mass (sum of accreted and in
  situ stars). Colours separate galaxies by B/T as given in the legend. Dashed
  lines of the same colour enclose 90 per cent of the distributions. Black
  solid and dashed lines correspond to the same quantities for the distribution
  shown in figure 5 (left panel) of Purcell {et~al.} (2007). Grey hatching and
  the black point with errorbars indicate likely values for the Milky Way
  ({Smith} {et~al.} 2007; {Li} \& {White} 2008; {Bell} {et~al.} 2008;
  {McMillan} 2011) and M31 ({Watkins} {et~al.} (2010); {Courteau} {et~al.}
  (2011) respectively.)}

  \label{fig:acc_mass_fraction}
\end{figure}

In the G11 model, lower galaxy formation efficiency above
$M_{\mathrm{peak}}$ is the result both of longer cooling times and of AGN
feedback \nocite{Kauffmann00,Croton06,Bower06}({Kauffmann} \& {Haehnelt} 2000;
{Croton} {et~al.} 2006; {Bower} {et~al.} 2006). This model follows the entire
hierarchy of galaxy formation and, with particle tagging, we have now computed
the dynamical evolution of all satellite disruption events in that model
directly from Millennium~II. These techniques mean that our model is
significantly more accurate and detailed than the empirical $z=0$ scaling
relations used by \nocite{Purcell07}{Purcell} {et~al.} (2007),
so it is useful to revisit their analysis.  Fig.~\ref{fig:acc_mass_fraction}
presents the data from Fig.~\ref{fig:acc_mass} in the same way as figure 4 of
\nocite{Purcell07}{Purcell} {et~al.} (2007), showing the ratio of
$M_{\mathrm{acc}}$ to $M_{\star}$ as a function of $M_{200}$. As expected, our
results are qualitatively similar\footnote{\nocite{Purcell07}{Purcell} {et~al.}
(2007) define their virial quantities at an overdensity of $\Delta=337$ rather
than $\Delta=200$ as we do. We have not corrected their results for this
difference in Fig.~\ref{fig:acc_mass_fraction}; doing so would shift their
curves to the right (higher virial mass) by $\lesssim0.1$~dex for plausible NFW
concentrations.} to \nocite{Purcell07}{Purcell} {et~al.}. However, our model
predicts a significantly steeper relation: an average accreted stellar mass
fraction of 30 per cent is reached in haloes with
$M_{200}\sim10^{12.5}\mathrm{M_{\odot}}$ rather than
$M_{200}\sim10^{13.5}\mathrm{M_{\odot}}$. 

\makebold{Both models appear roughly consistent with the (limited) data on late
type galaxies in haloes of $M_{200}\sim 10^{12} \mathrm{M_{\odot}}$,
represented in Fig.~\ref{fig:acc_mass_fraction} by M31 (black point;
\nocite{Watkins10}{Watkins} {et~al.} 2010; \nocite{Courteau11}{Courteau}
{et~al.} 2011) and the Milky Way (grey box;   \nocite{Smith07}{Smith} {et~al.}
2007; \nocite{Bell08} {Bell} {et~al.} 2008; \nocite{Li08mwmass}{Li} \& {White}
2008). In order to compare $M_{\mathrm{acc}}$ in our results and in those of
\nocite{Purcell07}{Purcell} {et~al.} to these observations in
Fig.~\ref{fig:acc_mass_fraction}, we have assumed that all observed halo stars
are accreted while all disc and bulge stars formed in situ. If this assumption
is reasonable, Fig.~\ref{fig:acc_mass_fraction} may imply that the Milky Way
has a less massive stellar halo than the average for its likely halo mass, or
that the lower values of $M_{200}$ are preferred.
Fig.~\ref{fig:acc_mass_fraction} reinforces the conclusion of C10 that
considerable scatter is expected in $M_{\mathrm{acc}}$ for Milky Way--like
galaxies. C10 attributed this to scatter in the accretion time and mass of the
most massive progenitor satellite (in the Milky Way, for example, the
Sagittarius dwarf has contributed to the stellar halo, but the Large Magellanic
Cloud has not).}

Our model disagrees with \nocite{Purcell07}{Purcell} {et~al.} (2007)
at higher $M_{200}$, where most galaxies are ellipticals and even a lower limit
to the accreted component is hard to identify by decomposing the light profile.
Our model predicts that a late type galaxy in a halo of $10^{13}
\mathrm{M_{\odot}}$ should have $\sim30$ per cent of its total stellar mass in
an extended $n\sim3$--$4$ spheroid of accreted stars
(\nocite{Purcell07}{Purcell} {et~al.} predict $\sim10$ per cent). This can be
tested with deep images of nearby massive disc galaxies (including S0s) for
which the total mass is well constrained by the rotation curve.

Finally, we discuss the progenitor galaxies that contribute stars to the
accreted component in our models. A progenitor is defined as a galaxy that is
disrupted within the `main branch' of a halo merger tree  -- each progenitor
may have many progenitors of its own, but all these are grouped together in
this definition based on only the final level in the hierarchy.
\makebold{Following the definition given in section 4 of C10, we find the most
massive galaxies in our sample typically have $\sim10$ significant progenitors
to their accreted component \nocite{Laporte12_arxiv}(see also {Laporte}
{et~al.} 2012).  Ellipticals have a reasonably tight correlation between the
number of significant progenitors and $M_{200}$; the lowest mass ellipticals in
our sample typically result from a single major merger and thus have only one
significant progenitor. Late type galaxies have a larger scatter in the number
of significant progenitors of $M_{\mathrm{acc}}$.  They can be dominated by one
massive object, in agreement with the findings of C10, but we also find cases
with 10 or more significant progenitors.}

\begin{figure} \includegraphics[width=84mm, trim=0.2cm 0.2cm 0.2cm 0.2cm, clip=True]{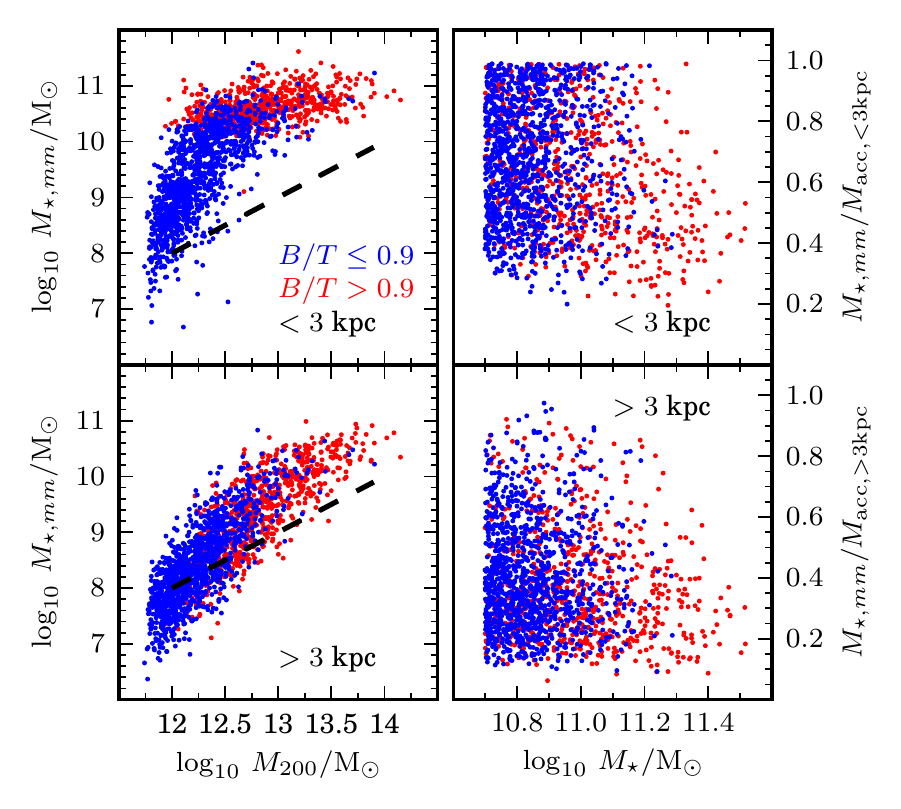}

  \caption{The stellar mass of the most massive progenitor, $M_{\star, mm}$, as
  a function of total halo mass $M_{200}$ (left) and the fraction of stellar
  mass acrreted from the most massive progenitor as a function of galaxy
  stellar mass $M_{\star}$ (right).  Upper and lower panels separate
  progenitors by the half mass radius $r_{50}$ of their stellar debris at
  $z=0$: $r_{50}<3$~kpc (the `bulge') and $r_{50}>3$~kpc (the `stellar halo'),
  respectively. The dashed black lines in the left-hand panels show the mass
  corresponding to $1\times10^{-4}M_{200}$.}

  \label{fig:mmmp}
\end{figure}

The left-hand column of Fig.~\ref{fig:mmmp} shows the mass of the most massive
progenitor in each of our galaxies. The upper and lower panels of this plot
separate two classes of progenitor, according to where their stars settle in
the main halo after they have been accreted. We define `bulge' progenitors to
be those that deposit more than half their stars within a radius of $3$~kpc
from the centre of the main halo at $z=0$. The rest are classified as `stellar
halo' progenitors (the same definition was used in C10). 

The most massive `bulge' progenitor is typically the most massive of all
accreted galaxies, which is not surprising because more massive satellites
suffer more dynamical friction and sink quickly to the centre of the potential.
Thus the relation between the mass of the most massive bulge progenitor and the
mass of the dark halo is similar to that shown in Fig.~\ref{fig:acc_mass}, with
a steep slope for late-type galaxies and an approximately constant value for
ellipticals. On the other hand, the mass of the most massive stellar halo
progenitor shows the same trend with $M_{200}$ in both late types and
ellipticals. Thus our model predicts the outer stellar haloes of all galaxies
to be equally diverse at fixed $M_{200}$. The most massive bulge progenitor is
$\sim10$ times as massive as the most significant contributor to the
halo up to $M_{200}\sim10^{13}\mathrm{M_{\sun}}$; at higher $M_{200}$ the most
significant bulge and stellar halo progenitors have similar mass.

The right-hand column of Fig.~\ref{fig:mmmp} shows the mass of the most massive
bulge and halo progenitors as a fraction of the total accreted stellar mass of
the main galaxy in the same region. The accreted bulges of late type galaxies
usually acquire at least $\sim40$ per cent of their stars from one progenitor,
reaching $\gtrsim 90$ per cent in many cases. On the other hand, the most
massive contributor to the accreted stellar halo typically accounts for no more
than $\sim30$~per~cent of its total mass, and rarely exceeds 50~per~cent. In
more massive galaxies ($M_{\star} \gtrsim 10^{11.2} \mathrm{M_{\sun}}$, which
mostly have G11 $B/T>0.9$) we find very few cases where more than 70~per~cent
of accreted bulge or halo stars originate in a single progenitor. As discussed
above, progenitors with $M_{200} > M_{\mathrm{peak}}$ (the most efficient halo
mass for galaxy formation) have roughly the same stellar mass as the central
galaxy that accretes them. This means that massive ellipticals can
undergo mergers that are `minor' in terms of $M_{200}$ (and thus numerous) but
`major' in terms of accreted stellar mass.
  
\section{Conclusions}
\label{sec:conclusions}

We have used dark matter particles in the Millennium~II simulation as dynamical
tracers of stellar populations, in order to study the hierarchical assembly of
galactic structure in the CDM model. \makebold{We have constrained the free
parameter of our particle tagging method, $f_{\mathrm{mb}}$, by comparison to
the galaxy mass--size relation for late-type galaxies and find acceptable
values in the range $\mathrm 1\% < f_{\mathrm{mb}} < 5\%$. We have identified
several limitations in the treatment of the in situ component that could be
improved on in future work (see Section~\ref{sec:limitations}). In particular,
the approximation of a universal constant value of $f_{\mathrm{mb}}$ could be
removed by computing it from other quantities in our semi-analytic model.} 

Our most important conclusions can be summarised as follows:

\begin{enumerate} 

  \item The stellar mass surface density profiles of galaxies in our model are
    well described by the sum of two \Sersic{} models, corresponding to the
    separate contributions of stars formed in situ and stars accreted from
    other galaxies. 

  \item The surface density of in situ stars falls off more rapidly
    with radius than the accreted component. In situ stars only make a
    significant contribution to surface brightness profiles out to
    $R\sim10$~kpc in $M_{200}\sim10^{12}\mathrm{M_{\sun}}$ halos or $40$~kpc in
    $M_{200}\sim10^{14}\mathrm{M_{\sun}}$ haloes for
    $f_{\mathrm{mb}}\sim1\%$.

  \item The outer isophotes of all massive galaxies are dominated by 
    accreted stars, which extend to the virial radius in most systems.

  \item Stellar mass surface density profiles show very little scatter
    from galaxy to galaxy at fixed $M_{200}$, particularly in
    group-scale haloes ($M_{200} > 10^{13.5}\, \mathrm{M_{\odot}}$).
    This is the consequence of strong correlations between halo mass, central
    star formation efficiency and the galaxy progenitor mass function, and is a
    basic feature of galaxy formation in the CDM model \nocite{Purcell07}(e.g.
    {Purcell} {et~al.} 2007). Comparison at fixed $M_{\star}$ shows more
    scatter, because of scatter in the relationship between $M_{\star}$ and
    $M_{200}$.

  \item The stellar content of galaxies can be in situ dominated or accretion
    dominated.  Accretion-dominated galaxies have more extended profiles and
    higher \Sersic{} index than in situ-dominated galaxies. They have
    approximately `power law' profiles from $10$--$100$~kpc that show no clear
    inflection at the transition between in situ and accreted components and
    are associated with haloes that host elliptical galaxies, according to the
    criteria described by G11. These massive haloes have usually been subject
    to `major' mergers (violent relaxation) after $z\sim1$, which reshape their
    in situ component.  They also suffer strong suppression of star formation
    by AGN feedback which prevents a compact core of in situ stars forming at
    low redshift.

  \item In situ dominated profiles are typical of less massive haloes (up to
    $M_{200}\sim10^{13}\mathrm{M_{\sun}}$). In situ stars have a compact
    exponential distribution by construction in our model. Extended
    ($R_{50}\sim10$~kpc) accreted spheroids begin to dominate only at
    $R\gtrsim10$~kpc, causing a clear inflection in the circularly-averaged
    surface brightness profile. Neither compact bulges nor compact
    exponential elliptical galaxies are created by accretion in our model. 

  \item The transition from in situ dominated profiles to accretion
    dominated profiles with increasing $M_{200}$ is less clear when studied
    in terms of central stellar mass $M_{\star}$, because of scatter in the
    $M_{\star}$--$M_{200}$ relation in our model.

  \item \makebold{The central galaxies of massive groups and clusters typically
      have $\sim10$ accreted progenitors with similar stellar masses. In
      contrast, the bulges of less massive haloes are dominated by one
      progenitor, typically the most massive of those accreted; their stellar
      haloes are more diverse. The scatter in the number of significant
      progenitors of bulges and stellar haloes is larger at lower $M_{200}$.
      The progenitor mix is not significantly different in the stellar haloes
      of late and early type galaxies in our model. All these findings are
      readily explained by trends in star formation efficiency with halo mass
      in the $\Lambda$CDM model.}

  \item We have compared the results of our model with data from the literature,
   including deep surface brightness profiles of individual galaxies and stacks
   of LRGs and BCGs. Subject to the crude way in which we have assigned
   galaxies to halo mass bins and our simplification of a constant
   mass-to-light ratio, the range of shapes, amplitudes and scales seen in our
   simulated surface density profiles for accreted stars agree qualitatively
   with these observations.

  \item \makebold{We have stacked SDSS images of galaxies in bins of stellar mass
    to obtain average surface density profiles that can be compared directly to
    the results of our simulations. We find a weak trend of surface density
    amplitude with $M_{\star}$ but no clear change in profile shape. We find
    this is also the case in our simulations, once we account for the
    well-known bias in stellar mass measurements for massive galaxies due to
    the finite depth of SDSS photometry.}

\end{enumerate}

The next generation of deep sky surveys (culminating in LSST) will reach the
low surface brightness limits required to detect stellar haloes and accretion
remnants around the majority of low-redshift $L_{\star}$ galaxies. The dawn of
this new era in the observation of galaxy structure is a strong motivation for
the further study of these regions in simulations. We have shown that the
particle tagging technique is a straightforward and well-constrained extension
of the semi-analytic method with a number of interesting applications that
merit further investigation. These include stellar population gradients in
early-type galaxies; the kinematics of diffuse light; the effects of
interactions on the structure of satellite galaxies and environmental trends;
the frequency of tidal features and their correlations with other galaxy
properties; and the intracluster light of massive clusters.

\section*{Acknowledgments}

We thank the anonymous referee for their comments, which greatly improved the
clarity of the paper, and Stefano Zibetti, for providing data in electronic
form.  The Millennium~II Simulation databases used in this paper and the web
application providing online access to them were constructed as part of the
activities of the German Astrophysical Virtual Observatory. APC acknowledges a
National Natural Science Foundation of China cooperation and exchange grant,
no.  11250110509. CSF acknowledges an ERC Advanced Investigator grant (267291,
COSMIWAY). This work was supported in part by an STFC rolling grant to the
Institute for Computational Cosmology of Durham University. SW is grateful for
support from the ERC through an Advanced Investigator Grant (246797,
GALFORMOD). MB-K acknowledges support from the Southern California Center for
Galaxy Evolution, a multi-campus research program funded by the University of
California Office of Research. APC is grateful to the authors and maintainers
of the Python language core and its numerous open-source libraries (IPython,
NumPy, SciPy and PyTables) and in particular to the late John D. Hunter, author
of Matplotlib.

Funding for SDSS-III has been provided by the Alfred P. Sloan Foundation, the
Participating Institutions, the National Science Foundation, and the U.S.
Department of Energy Office of Science. The SDSS-III web site is
\url{http://www.sdss3.org/}.

SDSS-III is managed by the Astrophysical Research Consortium for the
Participating Institutions of the SDSS-III Collaboration including the
University of Arizona, the Brazilian Participation Group, Brookhaven National
Laboratory, University of Cambridge, Carnegie Mellon University, University of
Florida, the French Participation Group, the German Participation Group,
Harvard University, the Instituto de Astrofisica de Canarias, the Michigan
State/Notre Dame/JINA Participation Group, Johns Hopkins University, Lawrence
Berkeley National Laboratory, Max Planck Institute for Astrophysics, Max Planck
Institute for Extraterrestrial Physics, New Mexico State University, New York
University, Ohio State University, Pennsylvania State University, University of
Portsmouth, Princeton University, the Spanish Participation Group, University
of Tokyo, University of Utah, Vanderbilt University, University of Virginia,
University of Washington, and Yale University.

\appendix

\section{Differences with C10 and numerical convergence}
\label{appendix_b}

\begin{figure} \includegraphics[width=84mm, clip=True]{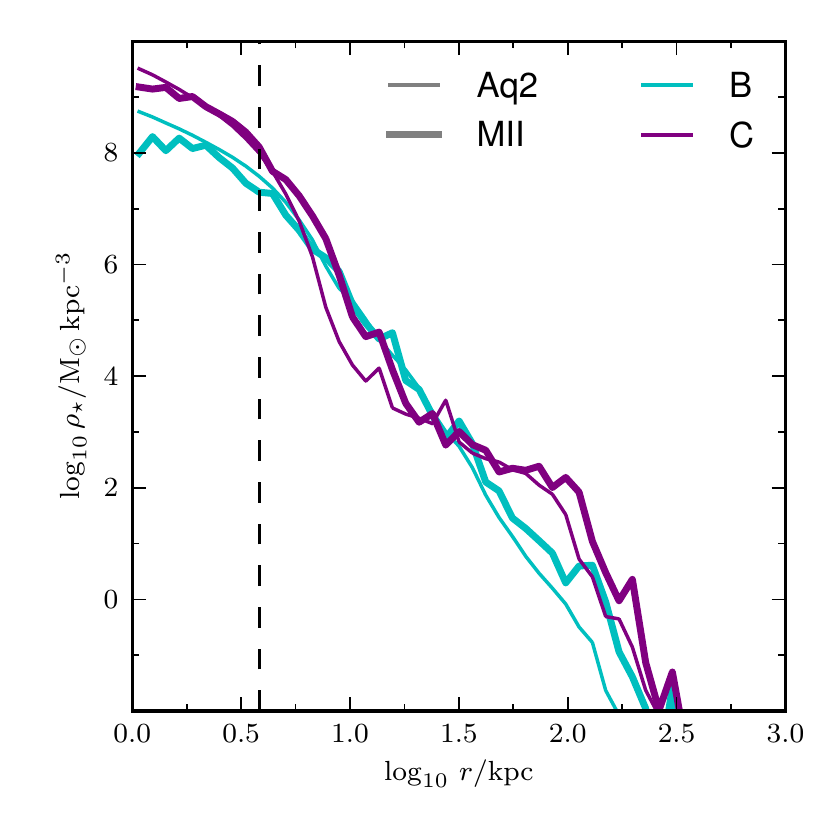}

  \caption{Comparison between particle-tagging stellar mass surface density
  profiles in Aquarius haloes B (cyan) and C (purple) at dark matter mass
  resolutions of $\sim1\times10^{3}\mathrm{M_{\odot}}$/particle (thin lines,
  Aquarius level 2) and $6.8\times10^{6}\mathrm{M_{\odot}}$/particle (thick
  lines, Millennium~II). The softening length of Millennium~II is shown by the
  vertical dashed line. These profiles differ from those shown by C10 because
  they are based on the G11 semi-analytic model and also include the in situ
  component.} 

  \label{fig:aquarius}
\end{figure}

There are a number of minor differences between the tagging procedure we use on
Millennium~II and that used on the Aquarius simulations by C10. First, as
discussed at length in the main text, we now tag particles to represent in situ
stars in our 1872 target central haloes as well as those formed in their
progenitors and satellite galaxies. In C10, the distribution of in situ stars
in the main halo was not considered, but only because they were not the focus
of that paper and excluding them made the computation less demanding.  

Second, the semi-analytic component of our model must deal with galaxies in
dark matter haloes that fall below the mass resolution limit of the simulation
because of tidal stripping.  We do this in the same way as G11, by estimating
the time for inspiral of the satellite due to dynamical friction in its parent
halo, and merging it with the central galaxy after that time. The satellite can
also be disrupted sooner if its density falls below that of the dark halo at
the satellite's pericentre. This approach differs from C10, where galaxies in
unresolved subhalos were merged instantly with the main galaxy in their parent
halo (in which case the disruption of the N-body subhalo is equivalent to the
disruption of the galaxy, as it would be in an SPH simulation with
collisionless star particles). C10 did this because of the high resolution of
their Aquarius simulations, which meant that it made almost no difference to
their results. However, at the lower resolution of Millennium~II, allowing
galaxies to survive after the disruption of their N-Body subhalos is necessary
for convergence of the semi-analytic model and its agreement with observations
(details are given in G11; see also \nocite{Font11}{Font} {et~al.} 2011). Therefore, our tagging
method must make allowance for new stars forming in galaxies with unresolved
haloes. We do this by tagging those stars to a single particle, the most-bound
particle of the halo at the time it was last resolved. Because very few stars
form in such haloes, this makes no practical difference. 

There is a more significant issue related to the treatment of semi-analytic
galaxies with unresolved haloes. Our analysis (for example, in constructing
density profiles) treats the stars in these galaxies as having been stripped,
because their dark matter particles are bound to the main dark matter halo.
This includes the most-bound particle to which we tag any residual star
formation.  Again, this treatment is similar to conventional hydrodynamic
simulations, which usually do not track galaxies below the resolution limit.
However, because of this, our definitions of stellar mass are not consistent
between the semi-analytic model (where these all stars belong to a galaxy) and
our tagged-particle model (where they are all stripped). These ambiguous stars
are easy to identify in the model, and we have verified that this choice makes
no significant difference to our results.

A related problem is that the lower resolution of Millennium~II (relative to
the Aquarius simulations of C10) makes {\em all} satellite galaxies somewhat
easier to disrupt (at fixed mass), because their density is artificially
reduced at radii below the force softening scale. Finally, it may be the case
that Millennium~II does not resolve a significant number of faint progenitors
at all, particularly for the least massive main galaxies in our sample.
However, as the much higher resolution simulations of C10 have shown, most of
the stellar mass accreted by Milky Way mass haloes comes from a small number of
their most massive progenitors, which are well resolved in Millennium~II. In
Fig.~\ref{fig:aquarius} we compare the high-resolution Aquarius simulations of
C10 with the same haloes simulated at the resolution of Millennium~II.
We find that resulting stellar mass density profiles have converged above the
softening scale of Millennium~II. The details of profile shape that distinguish
different Aquarius haloes in C10 are reproduced in Millennium~II, which has a
particle mass almost four orders of magnitude larger. This is the case
for both the accreted and in situ components.  This implies that the resolution
effects described above are not very important for our results (and notably,
that the underlying semi-analytic model of G11 has also converged with regard
to the predicted stellar mass of the Aquarius progenitors).

The final difference with C10 is that we do not postpone the tagging process to
later snapshots in cases where the target halo is deemed to be out of
equilibrium.  Although well-motivated, C10 found that this procedure makes the
implementation of the method much more complex but has little influence on the
outcome\footnote{At the level of the other approximations in this method, it is
arguable that allowing assignments to non-equilibrium haloes might be a
reasonable representation of the messy nature of star formation in mergers.}.
Other technical subtleties, including tagging a fixed number of particles
rather than a fixed fraction in subhalos that are losing mass, are dealt with
as described in C10.

\section{SDSS stacking analysis}
\label{appendix_d}

In Section~\ref{sec:sdss} we stacked galaxies from SDSS DR9 for comparison with
our models. This appendix describes our method for selecting galaxies in SDSS
and stacking their images. We intend to explore a number of important
systematic uncertainties in more detail and present further results in a
separate paper (D'Souza et al. in preparation), so our analysis here should be
considered preliminary.

Our starting point is the MPA-JHU SDSS `value-added' catalogue, which provides
an estimate of stellar mass for galaxies with spectra in DR7 based on fitting
an SED to their \texttt{modelMag} photometry \nocite{Kauffmann03a,Salim07}({Kauffmann} {et~al.} 2003a; {Salim} {et~al.} 2007). From
this catalogue, we selected isolated central galaxies in the redshift range
$0.07<z<0.08$ by applying the criteria of \nocite{WangWenting12}{Wang} \& {White} (2012): a galaxy of
apparent $r$ band magnitude $m$ is considered isolated if there are no galaxies
in the spectroscopic catalogue at a projected radius $R < 0.5\, \mathrm{Mpc}$
and velocity offset $|\delta z| < 1000 \, \mathrm{km\,s}^{\, −-1 }$ with magnitude $m' <
m + 1$, and none within $R < 1 \, \mathrm{Mpc}$ and $|\delta z| < 1000\,
\mathrm{km\,s}^{-1}$ with $m' < m$.  We make no selection on colour or
morphological type.

We created $1\, \mathrm{Mpc^{2}}$ mosaics in the $g$, $r$ and $i$ bands
centered on each galaxy in our sample using the `corrected' sky-subtracted
frames from the SDSS Data Release 9 image server and \texttt{SWarp}
\nocite{swarp02}({Bertin} {et~al.} 2002). These three mosaics were stacked together to make a `master
image' from which a `master mask' was obtained using SExtractor
\nocite{sextractor96}({Bertin} \& {Arnouts} 1996). Other galaxies in the field were conservatively masked by
convolving the master image with an $8\times8$ pixel top hat kernel before
running \texttt{SExtractor}. The master mask was applied to each individual
mosaic. The masked mosaics were then transformed to $z=0.08$ with the
flux-conserving IRAF task \texttt{geotran}, cropped to a uniform size of
$1200\times1200$~pixels ($690\times690$~kpc at $z\sim0.08$) and corrected for
extinction following \nocite{Schlegel98}{Schlegel}, {Finkbeiner} \&  {Davis} (1998).

We assume that the sky subtraction provided by the SDSS imaging pipeline is
adequate for our analysis. The pipeline sky subtraction was known to have
shortcomings in early data releases, but was revised in DR8\footnote{See
\url{www.sdss3.org/dr9/imaging/images.php} and
\url{www.sdss3.org/dr9/imaging/caveats.php}} to improve the photometry of
extended low-surface brightness regions around low redshift galaxies
\nocite{Blanton11}({Blanton} {et~al.} 2011). We confirmed the quality of the sky subtraction in the DR9
images by carrying out tests using the earlier DR7 images with our own
background subtraction, and also by comparing DR9 surface brightness profiles
for galaxies in the Virgo cluster with the results of \nocite{Kormendy09}{Kormendy} {et~al.} (2009).

We binned our SDSS sample for stacking in mass bins as we did our simulations
in Fig.~\ref{fig:density_profiles_obs}, using the mode of the MPA-JHU mass PDF
corrected to a Hubble parameter $h=0.73$ for consistency with the cosmology of
our simulations\footnote{We have not accounted for a possible overestimate of
the DR7 \texttt{modelMag} magnitudes owing to uncertainties in the background
subtraction of extended galaxies pre-DR8.}.  This resulted in $\sim600$--$1200$
SDSS galaxies per bin. Mosaics in the $gri$ bands were stacked separately using
IRAF \texttt{imcombine}, taking the mean value\footnote{Median and mode
stacking produced very similar results.} of each pixel after clipping at the
10th and 90th percentiles.  Images were centered before stacking but were not
axis-aligned. We also removed a small residual background from each stack such
that the stellar mass surface density falls to zero at the periphery of the
mosaic.  The $g$, $r$ and $i$ band mosaics for each galaxy were not included in
the corresponding stacks if they included stars brighter than $r=12$, if they
fell in the upper $10^{th}$ percentile of the distribution of pixel RMS for
each mosaic in their mass bin, if they had more than $75\%$ of their central 60
arcseconds masked, or if the masking algorithm failed due to a crowded field.

In a given mass bin, the azimuthally averaged mass profile was derived from the
$r$ band stack assuming a constant mass-to-light ratio. The $M/L_{r}$ assumed
for each bin is the average of $M/L$ for the galaxies in the bin, derived from
their MPA-JHU mass and $r$ band \texttt{modelMag}; in order of $M_{\star}$ for
the four bins, $M/L_{r} = [2.154,2.240,2.391,2.516]$. We also used the $g$,
$r$ and $i$ stacks to derive a radially varying $M/L_{r}$ profile based on the
colour of the light in each annulus, using the prescriptions of \nocite{Bell03}{Bell} {et~al.} (2003)
with a Chabrier IMF. This only makes a significant difference in the outermost
radial bins of each profile. However, the potential for large colour errors at
faint magnitudes (particularly those involving the $i$ band) adds considerably
to the uncertainty in these radially varying profiles.

We have not carried out deconvolution of the PSF at any stage of our analysis.
Galaxies in the mass and redshift range of our sample are typically well
resolved, so the PSF mostly affects the very inner part of the light profile.
We have compared the profiles of individual galaxies in our stacks with similar
galaxies at lower redshift, with SDSS \texttt{profMean} profiles and with deep
data from the Virgo cluster \nocite{Kormendy09}({Kormendy} {et~al.} 2009) to verify that the PSF only has
a significant effect (very roughly, of the order of $\gtrsim0.1$ magnitudes per
square arcsecond in the surface brightness profile) at $R\lesssim
5~\mathrm{kpc}$. Nevertheless, particularly for the $i$ band magnitude and the
measurement of colours, PSF effects are known to be significant at much larger
radii \nocite{deJong08}({de Jong} 2008) and a complete analysis should include a more thorough
quantification of the PSF \nocite{Tal11}(e.g. {Tal} \& {van Dokkum} 2011).

We estimated the uncertainty in each annulus in the constant $M/L$ case as the
sum in quadrature of Poisson error in the flux per contributing pixel and the
average RMS pixel-by-pixel deviation of each image from the stack in that
annulus, the latter term accounting for the sample variance. A bootstrap
estimate of variance would be preferable to understand the effects of sample
variance, but proved to be computationally expensive.  Bootstrapping on
sub-samples of 100 galaxies suggested that our algorithm underestimates the
variance in the outer regions of the profile by a factor of 2, so we multiply
our variance estimate by $2$ to obtain the error bar in
Fig.\ref{fig:density_profiles_sdss}. This crude estimate of uncertainty is
sufficient to indicate the largest radius to which each stacked profile is
robust, but it could be made substantially more accurate with further work.

In our selection of SDSS galaxies for stacking, we have used the MPA-JHU
stellar mass estimates. These masses are derived from the galaxy
\texttt{modelMag} magnitude, which attempts to correct for undetected light in
regions of low surface brightness by fitting the observed surface brightness
profile to one of two analytic models (exponential and $r^{1/4}$) and adopting
the total magnitude of the best fit. However, because the fits are to truncated
light profiles (owing to the finite surface brightness limit, of more
importance for more extended galaxies) and assume a fixed \Sersic{} index $n=4$
for the early-type model, they are likely to suffer from a bias relative to the
true total light similar to that of the Petrosian magnitude \texttt{petroMag}.
In theory \texttt{petroMag} is a substantial underestimate of the total light
for galaxies with \Sersic{} index $n \gtrsim 4$, but includes almost all of the
light for galaxies with $n\sim1$ \nocite{Graham05_petro, Lauer07,
Blanton11}({Graham} {et~al.} 2005; {Lauer} {et~al.} 2007; {Blanton} {et~al.}
2011).

The bias in \texttt{modelMag} is more difficult to reproduce in our simulations
because it depends on the fitting process itself (for example, in the behaviour
of fits to multi-component galaxies, or with substantial noise).  Furthermore,
we find that the median offset between \texttt{modelMag} and \texttt{petroMag}
for galaxies in our SDSS sample is generally small ($\sim0.1$ magnitudes). We
therefore compute the `Petrosian' mass from the simulations and use this as a
proxy for the SDSS \texttt{modelMag} mass in
Fig.~\ref{fig:density_profiles_sdss}. We do so with an algorithm analogous to
that used for Petrosian flux reported by SDSS (i.e.  the mass within $2r_{p}$,
using the same definition of $r_{p}$). The resulting $M_{\mathrm{pet}}$
underestimates $M_{\star}$ for the most massive galaxies, thereby
redistributing galaxies with high-$n$ profiles to lower mass bins in our
simulated stacks.

\bibliography{}
\bsp

\label{lastpage}

\end{document}